\documentclass[oldversion]{aa}  
\usepackage{graphicx,amsmath}
\usepackage{amssymb}
\usepackage{color}
\usepackage{natbib}             
\usepackage{url}
\usepackage{multirow}
\bibpunct{(}{)}{;}{a}{}{,} 

\def\approxinf{%
  \def\p{%
    \setbox0=\vbox{\hbox{$<$}}%
    \ht0=0.6ex \box0 }%
  \def\s{%
    \vbox{\hbox{$\sim$}}%
  }%
  \mathrel{\raisebox{0.7ex}{%
      \mbox{$\underset{\s}{\p}$}%
    }}%
}

\def\approxsup{%
  \def\p{%
    \setbox0=\vbox{\hbox{$>$}}%
    \ht0=0.6ex \box0 }%
  \def\s{%
    \vbox{\hbox{$\sim$}}%
  }%
  \mathrel{\raisebox{0.7ex}{%
      \mbox{$\underset{\s}{\p}$}%
    }}%
}

\begin{document}

   \title{An evaporating planet in the wind: \\
   stellar wind interactions with the radiatively braked exosphere of GJ\,436 b}
                                   
   \author{
   V.~Bourrier\inst{1},
   A.~Lecavelier des Etangs\inst{2},
   D.~Ehrenreich\inst{1},
   Y.~A.~Tanaka\inst{3,4},\and
   A.~A.~Vidotto\inst{1,5}}
   
\authorrunning{V.~Bourrier et al.}
\titlerunning{Comprehensive study of the upper atmosphere and stellar environment of GJ 436 b}

\offprints{V.B. (\email{vincent.bourrier@unige.ch})}

\institute{
Observatoire de l'Universit\'e de Gen\`eve, 51 chemin des Maillettes, 1290 Sauverny, Switzerland\and 
Institut d'astrophysique de Paris, UMR7095 CNRS, Universit\'e Pierre \& Marie Curie, 98bis boulevard Arago, 75014 Paris, France 
\and 
Department of Physics, Nagoya University, Nagoya, Aichi 464-8602, Japan
\and 
National Astronomical Observatory of Japan, Mitaka, Tokyo 181-8588, Japan
\and 
School of Physics, Trinity College Dublin, The University of Dublin, Dublin-2, Ireland
}
   
   \date{} 

  \abstract
{Observations of the warm Neptune GJ\,436b were performed with HST/STIS at three different epochs (2012, 2013, 2014) in the stellar Lyman-$\alpha$ line. They showed deep, repeated transits that were attributed to a giant exosphere of neutral hydrogen. The low radiation pressure from the M-dwarf host star was shown to play a major role in the dynamics of the escaping gas and its dispersion within a large volume around the planet. Yet by itself it cannot explain the specific time-variable spectral features detected in each transit. Here we investigate the combined role of radiative braking and stellar wind interactions using numerical simulations with the EVaporating Exoplanet code (EVE) and we derive atmospheric and stellar properties through the direct comparison of simulated and observed spectra.  \\
The first epoch of observations is difficult to interpret because of the lack of out-of-transit data. In contrast, the results of our simulations match the observations obtained in 2013 and 2014
well. The sharp early ingresses observed in 2013 and 2014 come from the abrasion of the planetary coma by the stellar wind. Spectra observed at later times during the transit can be produced by a dual exosphere of planetary neutrals (escaped from the upper atmosphere of the planet) and neutralized protons (created by charge-exchange with the stellar wind). We find similar properties at both epochs for the planetary escape rate ($\sim$2.5$\times$10$^{8}$\,g\,s$^{-1}$), the stellar photoionization rate ($\sim$2$\times$10$^{-5}$s$^{-1}$), the stellar wind bulk velocity ($\sim$85\,km\,s$^{-1}$), and its kinetic dispersion velocity ($\sim$10\,km\,s$^{-1}$, corresponding to a kinetic temperature of 12\,000\,K). We also find high velocities for the escaping gas ($\sim$50 - 60\,km\,s$^{-1}$) that may indicate
magnetohydrodynamic (MHD) waves that dissipate in the upper atmosphere and drive the planetary outflow. In 2013 the high density of the stellar wind ($\sim$3$\times$10$^{3}$cm$^{-3}$) led to the formation of an exospheric tail that was mainly composed of neutralized protons and produced a stable absorption signature during and after the transit. \\
The observations of GJ\,436 b allow for the first time to clearly separate the contributions of radiation pressure and stellar wind and to probe the regions of the exosphere shaped by each mechanisms. The overall shape of the cloud, which is constant over time, is caused by the stability of the stellar emission and the planetary mass loss, while the local changes in the cloud structure can be interpreted as variations in the density of the stellar wind.}


   \maketitle

\section{Introduction}
\label{intro} 

\subsection{Atmospheric escape}

Transit observations in the Lyman-$\alpha$ line of neutral hydrogen led to the detection of extended exospheres around the hot Jupiters HD\,209458b (\citealt{VM2003,VM2004}) and HD\,189733b (\citealt{Lecav2010, Lecav2012}; \citealt{Bourrier2013}), the warm Neptune GJ\,436b (\citealt{Kulow2014}; \citealt{Ehrenreich2015}), and the warm Jupiter 55 Cnc b (\citealt{Ehrenreich2012}). The intense stellar X-ray and extreme ultraviolet energy input at the base of a hydrogen-rich thermosphere has been shown to be responsible for the expansion of the upper atmospheric layers (e.g., \citealt{Lammer2003}; \citealt{Lecav2004};  \citealt{Koskinen2013a,Koskinen2013b}). Heavier species can be carried to high altitudes through collisions with the expanding flow of hydrogen, and several metals and ions were detected around these planets (\citealt{VM2004}, \citealt{Linsky2010}; \citealt{VM2013}; \citealt{Ballester2015}; \citealt{BJ_ballester2013}; \citealt{Fossati2010}, \citealt{Haswell2012}), confirming that their atmospheres are in a state of hydrodynamic blow-off.\\
Transit observations at high resolution in the UV have also been used to probe the structure of these extended exospheres, revealing that they are shaped by interactions with the host star such as photoionization, radiation pressure, and stellar wind interactions (e.g., \citealt{Holmstrom2008}; \citealt{Ekenback2010}; \citealt{Bourrier_lecav2013}; \citealt{BJ_ballester2013}; \citealt{Bourrier2014}; \citealt{Kislyakova2014}; \citealt{Guo2016}; \citealt{Schneiter2016}). Theoretical studies based on hydrodynamical simulations have also studied the processes that can affect the planetary outflow, such as charge-exchange reactions (e.g., \citealt{Tremblin2012}; \citealt{Christie2016}) or interactions with the planetary magnetic field (e.g., \citealt{Khodachenko2015}). The interpretation of absorption signatures with 3D numerical models of atmospheric escape allows studying not only the properties of the planetary outflow, but also obtaining direct constraints on the star, such as its X/EUV emission and wind properties. The detection of temporal variability in the exosphere of the hot Jupiter HD\,189733b (\citealt{Lecav2012}) additionally underlined the importance of multi-epoch observations to study the evolution of these properties.\\
While hot Jupiters have been the focus of most studies, they are subject to moderate escape rates that only weakly affect their long-term evolution. By contrast, theoretical studies (e.g., \citealt{Lecav2004, Lecav2007}; \citealt{Ehrenreich_desert2011}; \citealt{Owen2012}; \citealt{Lopez2013}; \citealt{Kislyakova2014b}) and trends in the exoplanet population (e.g., \citealt{beauge2013}, \citealt{howard2012}) show that lower-density planets like mini-Neptunes or super-Earths with a large volatile envelope may be most significantly affected by evaporation, leading in the more extreme cases to a massive erosion of the atmosphere and the formation of rocky remnant cores. \\

\subsection{GJ\,436 b}
\label{GJ436_obs}

\begin{figure}  
\includegraphics[trim=0cm 0cm 1cm 0cm,clip=true,width=\columnwidth]{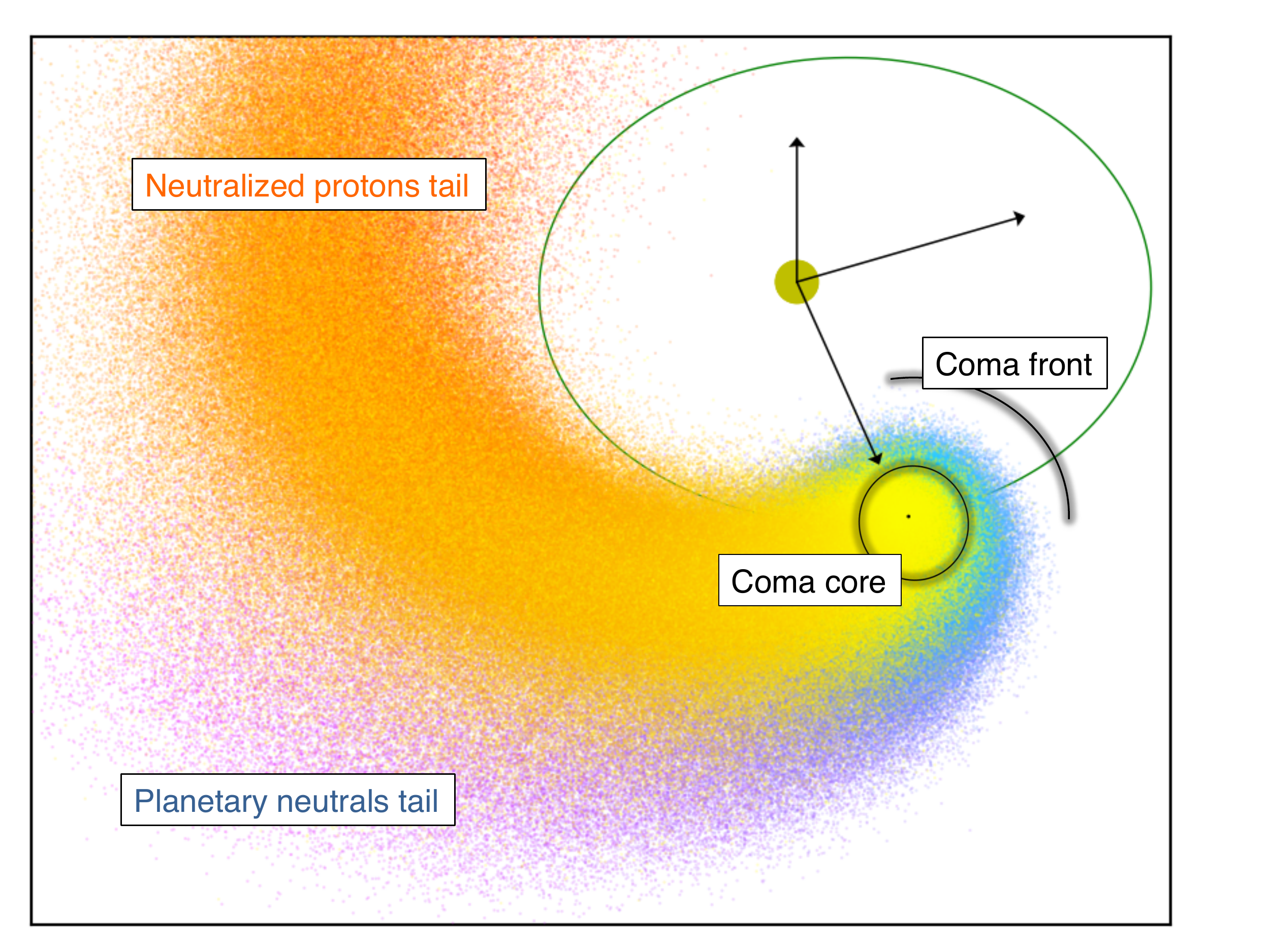}
\caption[]{Schematic representation of the neutral hydrogen cloud surrounding GJ\,436 b, displaying the different regions of the exosphere. Colors distinguish between the two populations of hydrogen atoms with different origins that compose the cloud.}
\label{cloud_schematics}
\end{figure}

Located at the edge of the sub-Jupiter desert (\citealt{beauge2013}), the warm Neptune GJ\,436 b ($R_{p}=4.2$\,$R_{Earth}$, $P=2.6$\,days, $a=0.0287$\,au; \citealt{Butler2004}; \citealt{Gillon2007}) is an ideal candidate to investigate the evaporation of low-mass gaseous planets. In contrast to other known evaporating planets orbiting G- and K-type stars, GJ\,436 b is hosted by an M dwarf with moderate irradiation ($M_\mathrm{\star}$ = 0.45$M_\mathrm{\sun}$, $R_\mathrm{\star}$ = 0.44$R_\mathrm{\sun}$), which enables us to study a new regime of atmospheric escape and star-planet interactions. The brightness of the host star ($V=10.7$) and its close proximity to Earth ($d=10.14$\,pc) makes GJ\,436 a good target for transit observations in the Lyman-$\alpha$ line. Using HST/STIS, \citet{Kulow2014} identified a deep absorption signature from neutral hydrogen after the end of the optical transit, but their interpretation was misled by an inaccurate transit ephemeris and by the lack of an out-of-transit reference for the flux in the stellar Lyman-$\alpha$ line (\citealt{Ehrenreich2015}). Using two additional HST observations, \citet{Ehrenreich2015} revealed a deeper signature repeated over the three epochs of observations, which shows that GJ\,436 b is surrounded by a giant coma of neutral hydrogen that is large enough to occult the stellar disk several hours before the optical transit, and that is trailed by a long cometary tail that could remain detectable for many hours after the optical transit (Fig~\ref{cloud_schematics}). \\

Following this detection, \citet{Bourrier2015_GJ436} studied the role played by stellar radiation pressure on the exosphere structure of GJ\,436 b and its transmission spectrum. In contrast to hot evaporating planets, radiation pressure from its M-dwarf host star is too low to overcome stellar gravity and repel exospheric hydrogen atoms from the star. But it is still high enough to brake the gravitational deviation of the atoms toward the star, allowing their dispersion within a large volume around the planet. This effect is referred to as radiative braking. While radiative braking explains the size of the coma and the blueshifted velocity range of the observed absorption signatures up to about -120\,km\,s$^{-1}$ well, it does not account for the variations in the depth and duration of the absorption signal at the different phases of the transits. Furthermore, even though the overall signature of the GJ\,436 b exosphere is very similar in the three different epochs, specific features to each visit cannot be explained by the radiation pressure, which is due to the Lyman-$\alpha$ line that was shown to be extremely stable over time (\citealt{Ehrenreich2015}; \citealt{Bourrier2015_GJ436}). Our goal in this paper is to investigate the coupled effects of stellar wind interactions and radiation pressure on the exosphere, using numerical simulations of the GJ\,436 system with the EVaporating Exoplanet code (EVE). We also compare simulated spectra with the observations in each epoch to measure the values of planetary and stellar parameters that shape the exosphere. \\
This study is based on the three existing transit data sets of GJ\,436 b observations (Table~\ref{tab:log}) taken in the H\,{\sc i} Lyman-$\alpha$ line with the Space Telescope Imaging Spectrograph (STIS) instrument onboard the Hubble Space Telescope (HST). These datasets are described in \citet{Bourrier2015_GJ436}. The code EVE was also described in this paper, and in Sect.~\ref{model} we summarize its main characteristics and detail the extension developed for stellar wind interactions. In Sect.~\ref{struct} we investigate the differences between radiation pressure and stellar wind on the spatial and velocity structure of the exosphere and its spectral signature. EVE simulations are compared to the observed spectra in Sect.~\ref{results} to measure the properties of GJ\,436 b environment at the different epochs of observations. These results are interpreted in Sect.~\ref{interpret}. In Sect.~\ref{sect:separation} we discuss how the effects of radiation pressure and stellar wind interactions can be disentangled, and we conclude in Sect.~\ref{sect:conclu}.\\


\section{Modeling the exosphere with the code \textit{EVE}}
\label{model}

EVE is a 3D numerical code developed to calculate the structure of an exoplanet upper atmosphere and its transmission spectra. The code was used in \citet{Ehrenreich2015} to perform a preliminary fit of the three combined observations of GJ\,436b neutral hydrogen exosphere. It was then used in \citet{Bourrier2015_GJ436} to investigate the influence of radiation pressure. In this paper, we implement a stellar wind extension to the code to study the effect of charge exchange on the exosphere and to measure the properties of GJ\,436b environment at the different observation epochs. The main physical parameters used for GJ\,436 b and its host star are given in Table~\ref{param_sys}.\\

\begin{table}[tbh]
\caption{Physical parameters for the GJ\,436 system.}                                                                           
\begin{tabular}{llcccc}
\hline
\hline
\noalign{\smallskip}
Parameters                                        & Symbol                                                                                                         & Value                                                                                                                                                                                               \\
\noalign{\smallskip}
\hline
\noalign{\smallskip}
Distance from Earth             & $D_{\mathrm{*}}$                                                                                &    10.14\,pc                                                                                                                                 \\
\noalign{\smallskip}
Star radius                                             & $R_{\mathrm{*}}$                                                                                &    0.44$\,R_{\mathrm{\sun}}$                                                                                 \\
\noalign{\smallskip}
Star mass                                                       & $M_{\mathrm{*}}$                                                                                  &    0.45$\,M_{\mathrm{\sun}}$                                                                                 \\
\noalign{\smallskip}
Planet radius           & $R_{\mathrm{p}}$                                                                               &    0.35$\,R_{\mathrm{Jup}}$                                                                   \\
\noalign{\smallskip}
Planet mass                             & $M_{\mathrm{p}}$                                                                                &    0.073$\,M_{\mathrm{Jup}}$                                                                         \\
\noalign{\smallskip}
Orbital period                          & $P_{\mathrm{p}}$                                                                               &    2.644$\,days$  \\
\noalign{\smallskip}
Transit center                          & $T_{\mathrm{0}}$                       &    2454865.083208$\,BJD$  \\
\noalign{\smallskip}
Semi-major axis                         & $a_{\mathrm{p}}$                                                                                &    0.0287$\,au$      \\
\noalign{\smallskip}
Eccentricity                            & $e$                                                                             &    0.16      \\
\noalign{\smallskip}
Argument of periastron  & $\omega$                                                                                &    327$^{\circ}$     \\

\noalign{\smallskip}
Inclination                                       & $i_{\mathrm{p}}$                                                                              &    86.7$^{\circ}$  \\
\noalign{\smallskip}
\hline
\hline
\end{tabular}
\label{param_sys}
\end{table}

\subsection{General description} 

A detailed description of the code EVE and the numerical settings used for GJ\,436 b can be found in \citet{Bourrier2015_GJ436}, and we summarize its main features here. The upper planetary atmosphere is divided into two different regimes that are joined at the mean altitude of the Roche lobe. The bottom layers of the atmosphere are described analytically, while Monte Carlo particle simulations are used to compute the dynamics of neutral hydrogen metaparticles in the upper atmospheric layers. Particles are subjected to the stellar and planetary gravities, the stellar radiation pressure, and the inertial force linked to the non-Galilean stellar reference frame. They are also affected by stellar photoionization and charge exchange with the stellar wind (Sect.~\ref{st_wind_model}). We use the term projected velocity to refer to the projection of a particle velocity on the star-Earth line of sight, and refer to its projection on the star-particle axis as radial velocity (\textit{i.e.}, the radial coordinate in the star reference frame). While we measure spectra as a function of projected velocity, radiation pressure is proportional to the flux in the intrinsic Lyman-$\alpha$ line and therefore it varies with the radial velocity of hydrogen atoms. The intrinsic stellar Lyman-$\alpha$ line can only be reconstructed by accounting for interstellar medium (ISM) absorption, and we used the lines reconstructed for Visits 2 and 3 in \citet{Bourrier2015_GJ436}. As shown by these authors, the lack of out-of-transit observations in Visit 1 prevents reconstructing the intrinsic line, and we used the line obtained for Visit 2 as a proxy. The density and velocity structures of the gas in the exosphere depend on three free model parameters: the escape rate of neutral hydrogen ($\dot{M}_{\mathrm{H^{0}}}$ in g\,s$^{-1}$, at the distance of the semi-major axis), the photoionization rate per atom ($\Gamma_{\mathrm{ion}}$ in s$^{-1}$, at the distance of the semi-major axis), and the velocity of the planetary wind at the Roche lobe ($v^{\mathrm{p}}_{\mathrm{wind}}$, in km\,s$^{-1}$). We accounted for the effect of the orbital eccentricity ($e=0.16$, \citealt{lanotte2014}) that causes the escape rate and photoionization rate to vary in time with the inverse distance to the star squared. The effect of radiation pressure and photoionization on neutral hydrogen particles was calculated taking self-shielding within the exospheric cloud into account.\\
Constraints on the model parameters come from the direct comparison between the STIS observations and theoretical Lyman-$\alpha$ line spectra calculated at each time step with EVE at a resolution $\Delta \lambda=$0.04\,\AA\, corresponding to $\Delta v=$10\,km\,s$^{-1}$ (about half the resolution of the STIS spectra at 1215.67\,\AA). These theoretical spectra are affected by the planetary occultation, the exospheric absorption (taking the bulk motion, thermal, and natural broadening of the hydrogen gas into account), the interstellar medium (ISM) absorption, and STIS line spread function (LSF). The merit function for a given visit is the sum of the $\chi^2$ yielded by the comparison of the observed spectra with the theoretical spectra averaged during the time window of each observation (see Table~\ref{tab:log}). Data obtained in time-tag mode was sliced into two exposures per HST orbit (\citealt{Ehrenreich2015}), yielding a good compromise between signal quality and time sampling of the simulations. The fits were calculated in the velocity ranges [-200 ; -25] and [40 ; 200] \,km\,s$^{-1}$ for Visit 1
and [-200 ; -40] and [20 ; 200] \,km\,s$^{-1}$ for Visits 2 and 3. The line core was excluded because of airglow contamination and ISM absorption, while high velocities in the wings of the line did not have enough signal. To summarize, the fits were performed on eight exposures per visit for a total of about 210 data points. 

\begin{table*}
\caption{Log of GJ\,436b transit observations.}
\centering
\begin{tabular}{lcccccc}
\hline
\hline
\noalign{\smallskip}    
Phase         & Out-of-transit & Ingress & Transit      &\multicolumn{2}{c}{Egress}     \\      
\noalign{\smallskip}
\hline
\noalign{\smallskip}
Visit 1 (December 2012)                 &   -             & [-01:55 ; -01:30]      &   [-00:43 ; 00:05]  &   [00:52 ; 01:41]  &   [02:28 ; 03:17] \\ 
\noalign{\smallskip} 
Visit 2 (June 2013)             &[-03:23 ; -02:55]& [-02:01 ; -01:27]      &   [-00:26 ; 00:09]  &   [01:10 ; 01:45]  &   - \\ 
\noalign{\smallskip}
Visit 3 (June 2014)             &[-03:26 ; -02:58]& [-02:00 ; -01:25]      &   [-00:24 ; 00:10]  &   [01:11 ; 01:46]  &   - \\
\noalign{\smallskip}
\hline
\hline
\multicolumn{7}{l}{Note: The different phases relate to the transit of the extended exosphere of GJ\,436 b. Throughout the paper, we refer}\\
\multicolumn{7}{l}{explicitely to the occultation caused by the planetary disk alone as the optical transit. Time is given in hours and minutes,}\\
\multicolumn{7}{l}{and counted from the center of the optical transit.} \\
\end{tabular}
\label{tab:log}
\end{table*}

\subsection{Extension: stellar wind interactions}
\label{st_wind_model}

A stellar wind proton may gain an electron from the interaction with a neutral hydrogen atom in the planetary exospheric outflow. With the usual assumption (\citealt{Lindsay2005}) that charge transfer collisions result in little deflection of the interacting proton and no significant change in its kinetic energy, we consider that the population of neutralized protons keeps the velocity distribution of the stellar wind. With regard to observations in the Lyman-$\alpha$ line, charge exchange replaces the contribution of an exospheric neutral atom by that of a neutralized proton at the same position but with different velocity properties. We therefore handle this process in EVE as an impulsion given to the neutral atom undergoing charge exchange, so that afterward it moves with the velocity of the interacting proton. In that way, stellar wind protons need not be treated as an independent particle population because we only need to know the probability $dP$ that a given neutral hydrogen atom is accelerated by a proton during a simulation time step \textit{dt}: 
\begin{equation}
dP = 1-\exp[- \sigma_{\mathrm{HH^{+}}}(\Delta V) \, \Delta V \, n_{\mathrm{H^{+}}}  \,  dt]
\label{eq:ENA}
,\end{equation}
with $n_{\mathrm{H^{+}}}$ the stellar wind proton density in the vicinity of the hydrogen atom and $\Delta V = ||  \vec{V_{\mathrm{H}}} - \vec{V_{\mathrm{H^{+}}}} \|$ the relative velocity between the neutral atom and the interacting proton. Compared to \citet{Bourrier2015_GJ436}, we reduced the time step $dt$ to 2.6\,min to better account for the fast dynamics of the stellar wind. The cross section of the interaction $\sigma_{HH^{+}}$ is energy dependent. For relative velocities in our simulations lower than 1\,000\,km\,s$^{-1}$ (energy lower than 5.2\,keV), the formula from \citet{Lindsay2005} can be approximated to 
\begin{equation}
\sigma_{HH^{+}} = 10^{-20}(10.61-1.062\,ln(\Delta V))^2
,\end{equation}
with $\sigma_{HH^{+}}$ in m$^2$, $\Delta V$ in km\,s$^{-1}$. The velocity of the interacting proton is taken from the Maxwellian speed distribution of the stellar wind, defined analytically through its radial bulk velocity $V^{\mathrm{st}}_{\mathrm{bulk-wind}}$ and kinetic temperature $T^{\mathrm{st}}_{\mathrm{wind}}$, which are assumed to vary little over the spatial extension of the exosphere. We note that this kinetic temperature corresponds to the Maxwellian velocity dispersion of the proton population $v^{\mathrm{st}}_{\mathrm{therm-wind}}$=$\sqrt{k\,T^{\mathrm{st}}_{\mathrm{wind}}/m_{\mathrm{H^{+}}}}$ around the bulk motion of the stellar wind. This bulk motion is associated with a different temperature, generally in the order of millions of kelvins, in the frame of the Parker theory (e.g., \citealt{Vidotto2010}).  \\

Outside of the exosphere, we assumed that the stellar wind density decreases as a function of the distance from the star $r$ according to a quadratic law. Within the exosphere, hydrogen atoms shield each other from the stellar protons in the same way as from stellar photons. The proton density $n_{\mathrm{H^{+}}}$ at the distance $r$ from the star decreases with the penetration depth \textit{$\Lambda$} into the atmosphere as
\begin{align}
&n_{\mathrm{H^{+}}}(r)= n^{\mathrm{st}}_{\mathrm{wind}} \, \left(\frac{a_{\mathrm{p}}}{r}\right)^2  \, \exp[-\tau_\mathrm{HH^{+}}(\Lambda)] \\
&\tau_\mathrm{HH^{+}}(\Lambda) = \int\limits_{\mathrm{0}}^{\mathrm{\Lambda}}{n_{\mathrm{H}}(\mu) \, \sigma_\mathrm{HH^{+}}(\mu) \,  \frac{\Delta V(\mu)}{V_{bulk-wind}} \, d\mu} \nonumber
\end{align} 
\label{eq:npr_SS}
with $n^{\mathrm{st}}_{\mathrm{wind}}$ the value of the proton density at the distance of the planet semi-major axis $a_{\mathrm{p}}$, and $n_{\mathrm{H}}$ the neutral hydrogen density at an intermediate penetration depth $\mu$. In the upper layers of the exosphere, each metaparticle contributes to the optical depth $\tau_\mathrm{HH^{+}}$ at its own velocity. To calculate the contribution of the lower atmosphere, we used the equations from \citet{Fahr_Bzowski2004} to integrate $\Delta V(\mu)$ over the velocity distribution of the neutral hydrogen gas, assumed to be a Maxwellian centered on the planet orbital velocity. We simplified these calculations by using the average velocity of the stellar wind $V^{\mathrm{st}}_{\mathrm{bulk-wind}}$.\\
Our description for the stellar wind adds three more free parameters to the model: the bulk velocity ($V^{\mathrm{st}}_{\mathrm{bulk-wind}}$ in km\,s$^{-1}$) and kinetic dispersion ($v^{\mathrm{st}}_{\mathrm{therm-wind}}$ in km\,s$^{-1}$) of the proton distribution, which are representative of the wind conditions at the location of the planet, and the proton density ($n^{\mathrm{st}}_{\mathrm{wind}}$ in cm$^{-3}$, given at the semi-major axis of the planet). \\


\section{Structure of the exosphere}
\label{struct}

\subsection{Radiative braking and stellar wind interactions}
\label{sect:dynam}

\begin{figure*}
\centering
\begin{minipage}[b]{\textwidth}   
\includegraphics[trim=0.1cm 12.5cm 1.2cm 0cm,clip=true,width=0.353\textwidth]{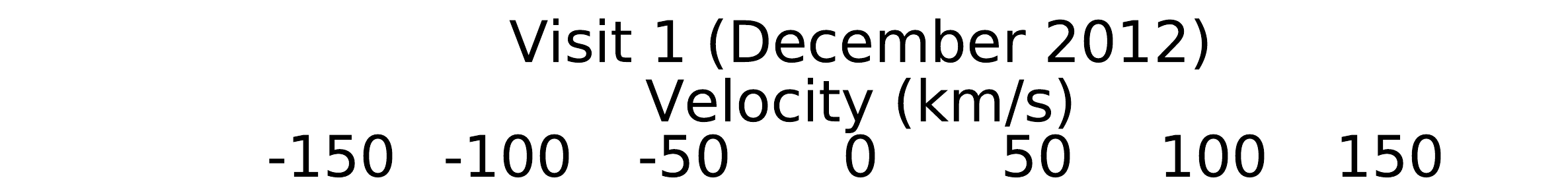}
\includegraphics[trim=2.5cm 0cm 1.2cm 0cm,clip=true,width=0.318\textwidth]{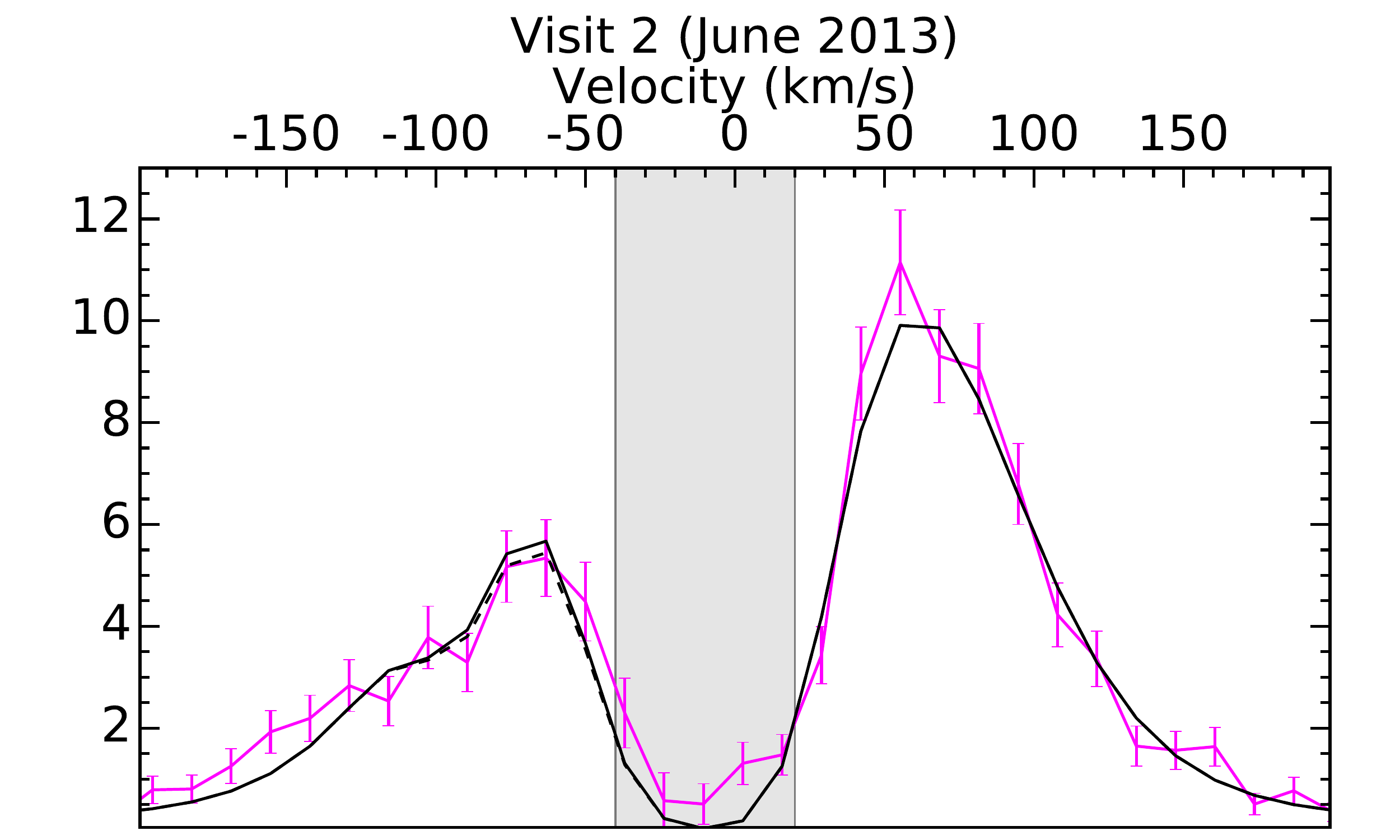}
\includegraphics[trim=2.5cm 0cm 1.2cm 0cm,clip=true,width=0.318\textwidth]{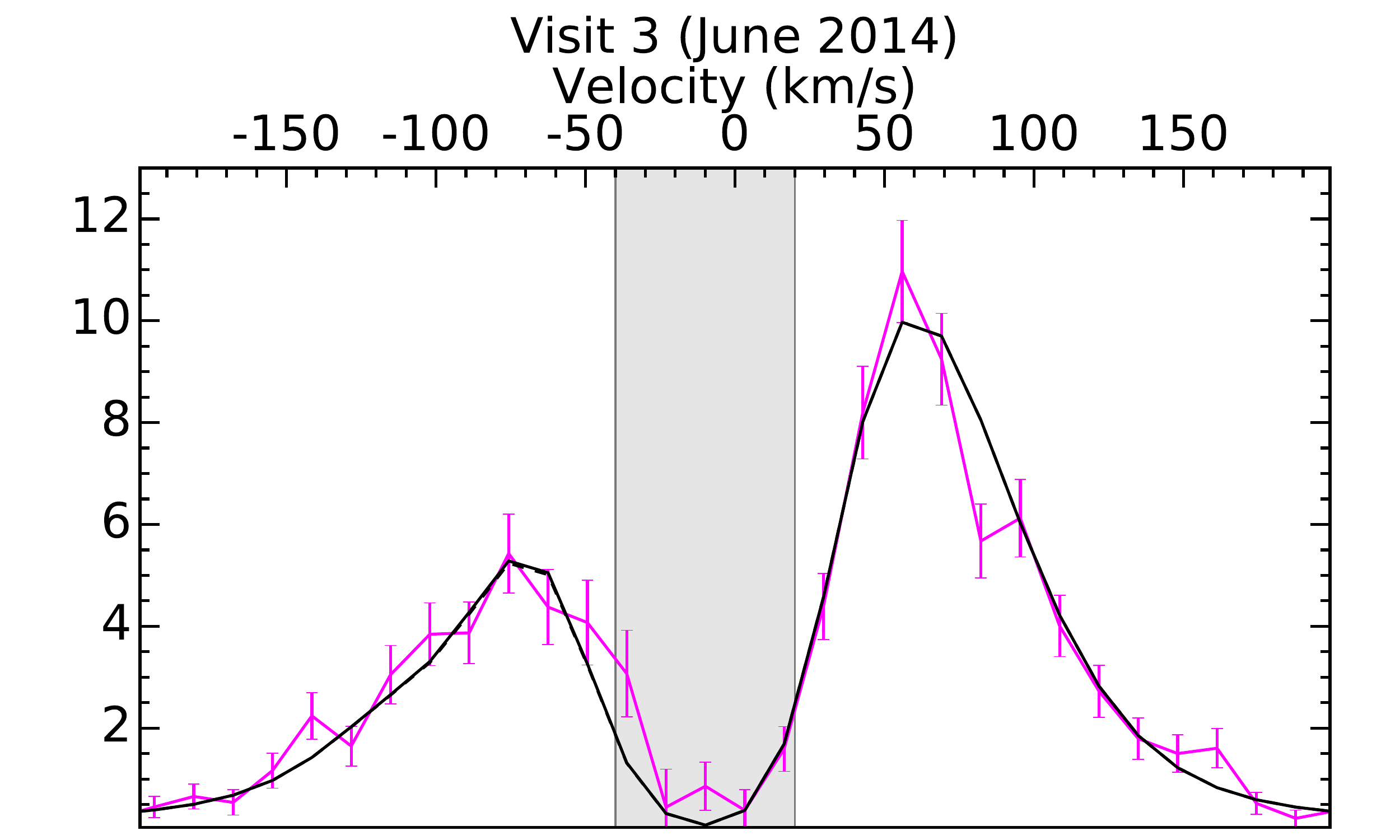}\\
\includegraphics[trim=0.1cm 0cm 1.2cm 2.5cm,clip=true,width=0.353\textwidth]{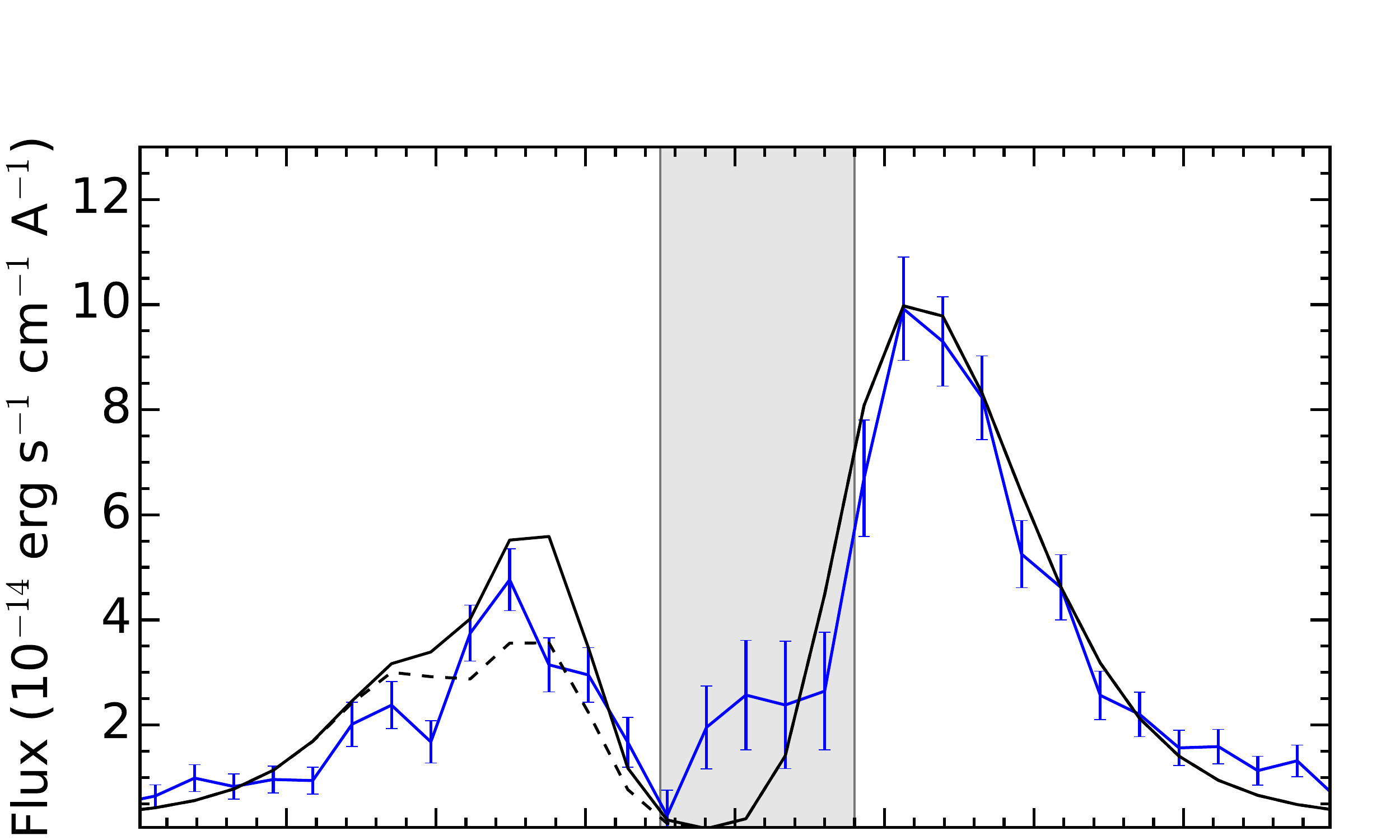}
\includegraphics[trim=2.5cm 0cm 1.2cm 2.5cm,clip=true,width=0.318\textwidth]{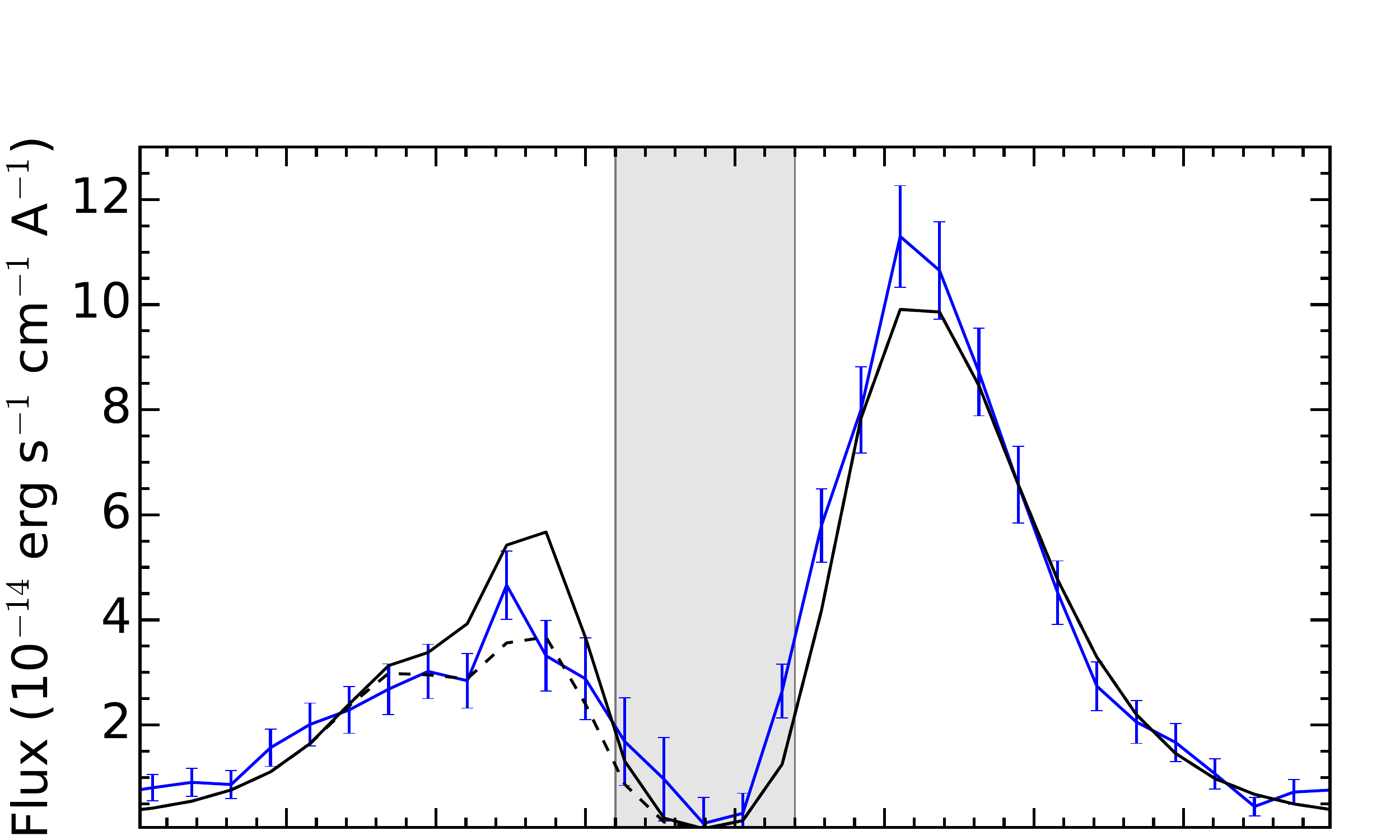}
\includegraphics[trim=2.5cm 0cm 1.2cm 2.5cm,clip=true,width=0.318\textwidth]{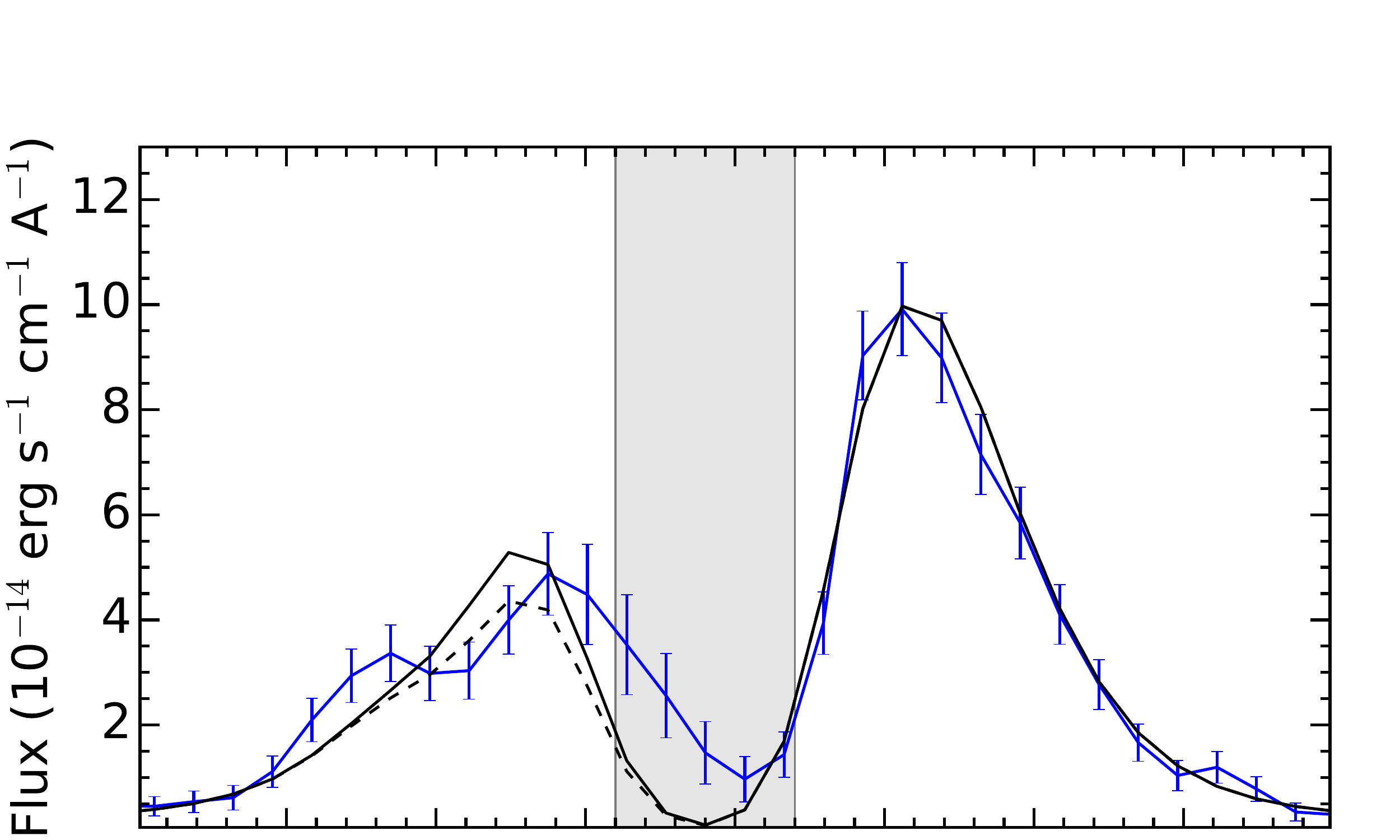}\\
\includegraphics[trim=0.1cm 0cm 1.2cm 2.5cm,clip=true,width=0.353\textwidth]{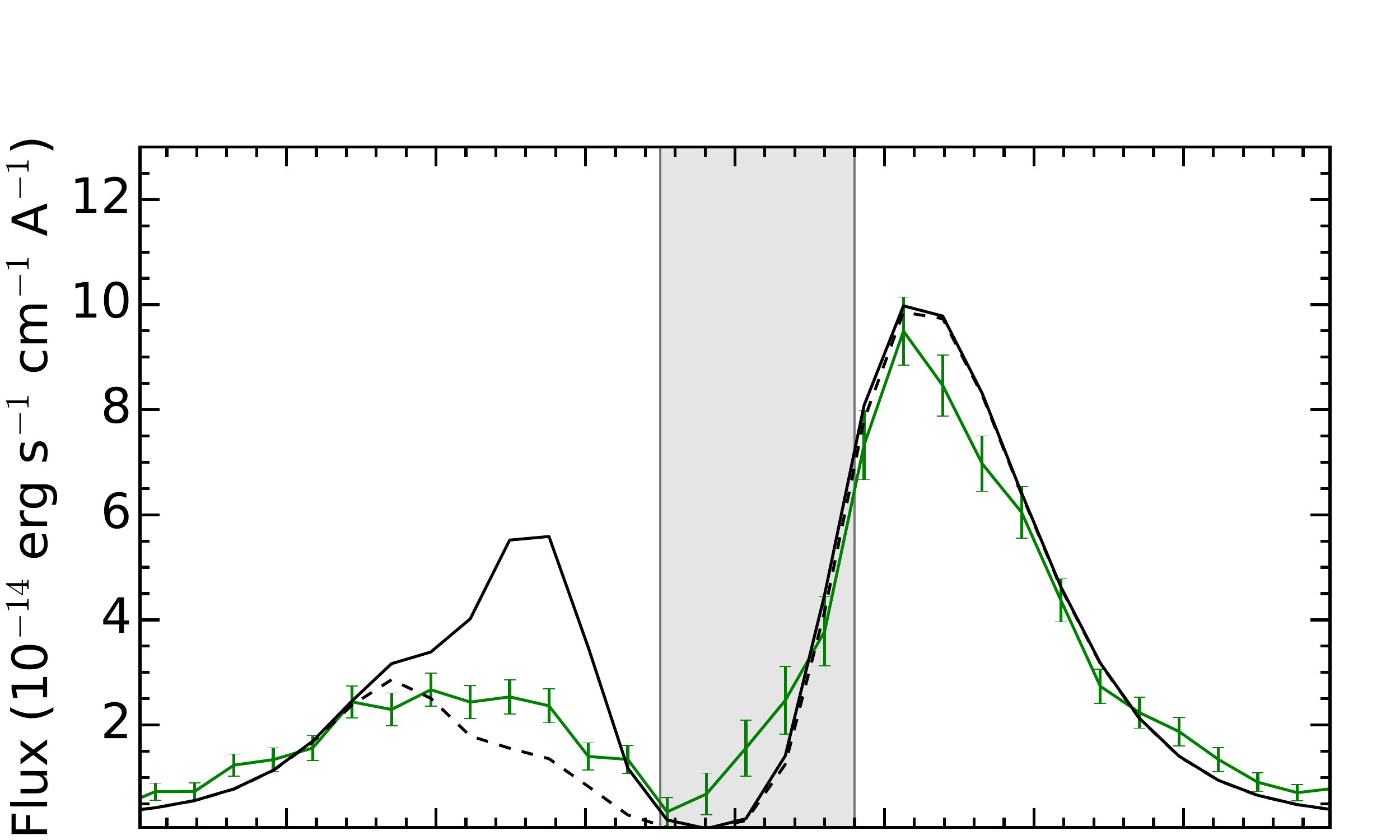}
\includegraphics[trim=2.5cm 0cm 1.2cm 2.5cm,clip=true,width=0.318\textwidth]{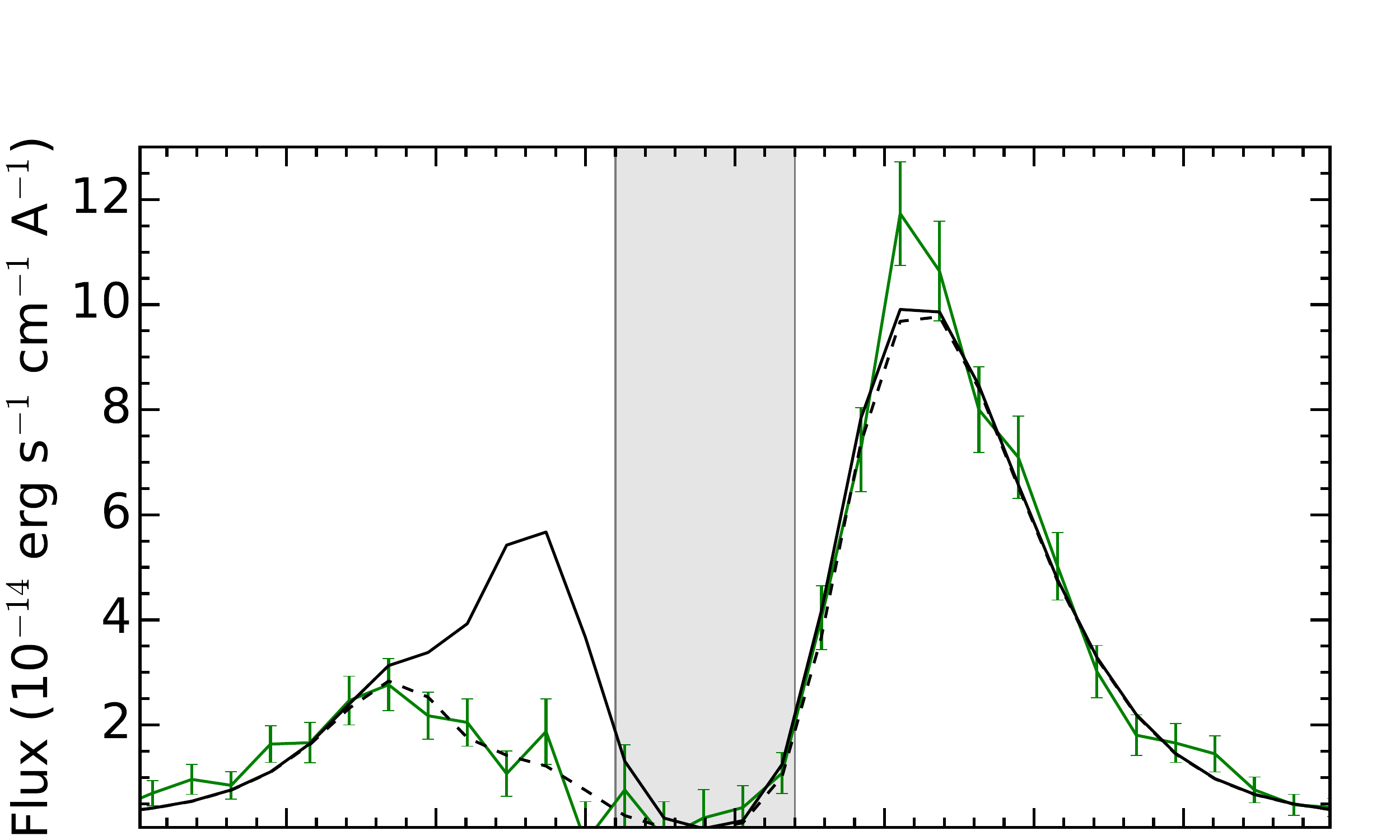}
\includegraphics[trim=2.5cm 0cm 1.2cm 2.5cm,clip=true,width=0.318\textwidth]{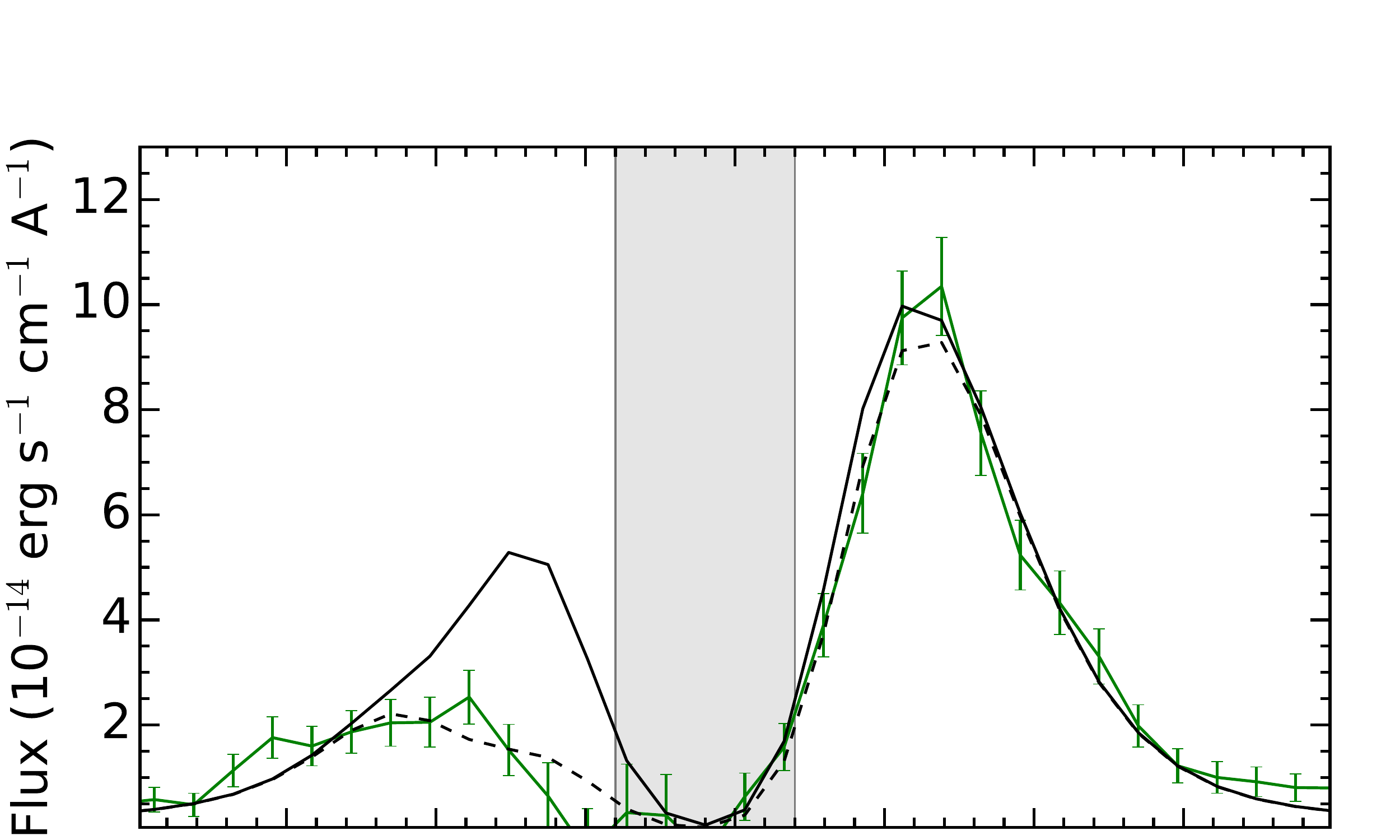}\\
\includegraphics[trim=0.1cm 0cm 1.2cm 2.5cm,clip=true,width=0.353\textwidth]{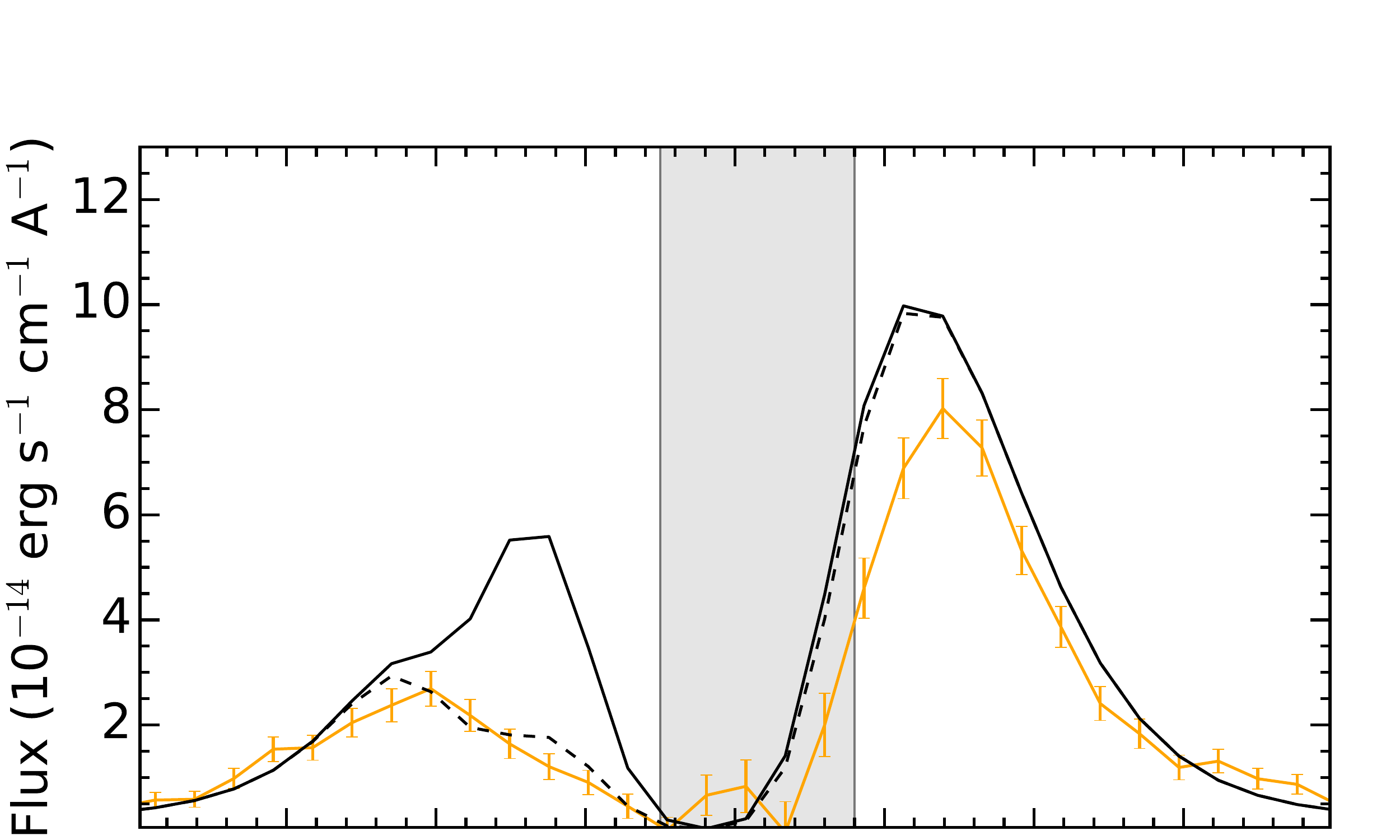}
\includegraphics[trim=2.5cm 0cm 1.2cm 2.5cm,clip=true,width=0.318\textwidth]{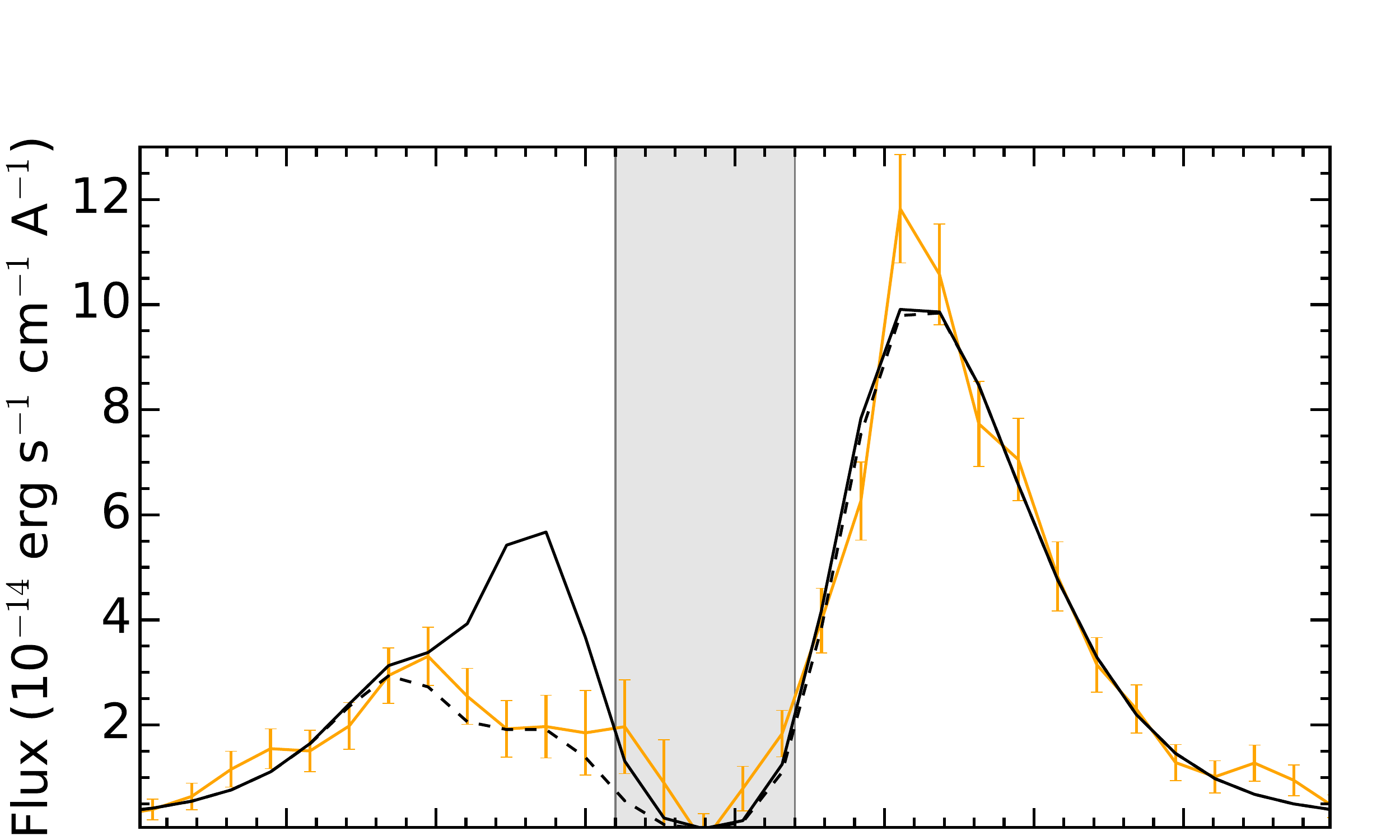}
\includegraphics[trim=2.5cm 0cm 1.2cm 2.5cm,clip=true,width=0.318\textwidth]{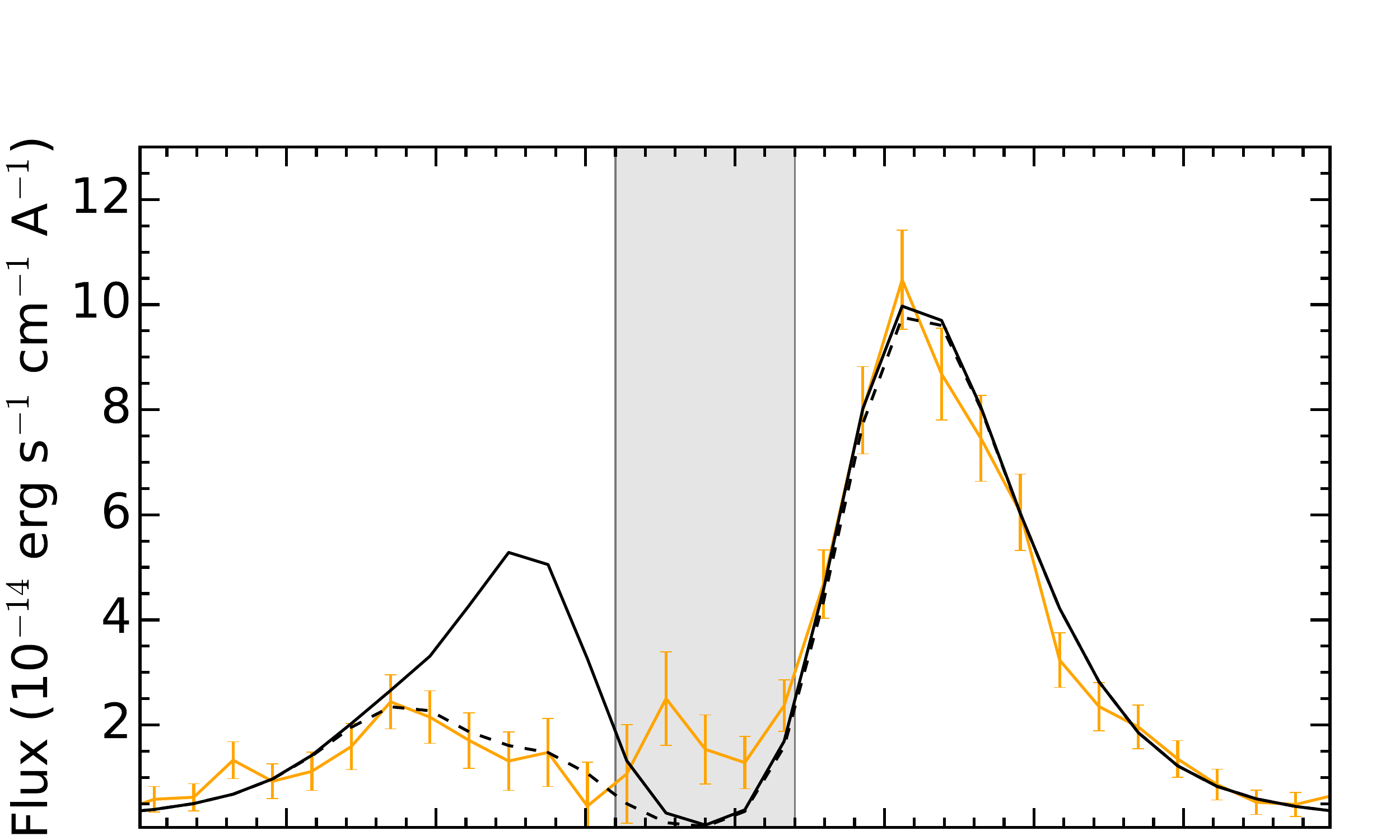}\\
\includegraphics[trim=0.1cm 0cm 1.2cm 2.5cm,clip=true,width=0.353\textwidth]{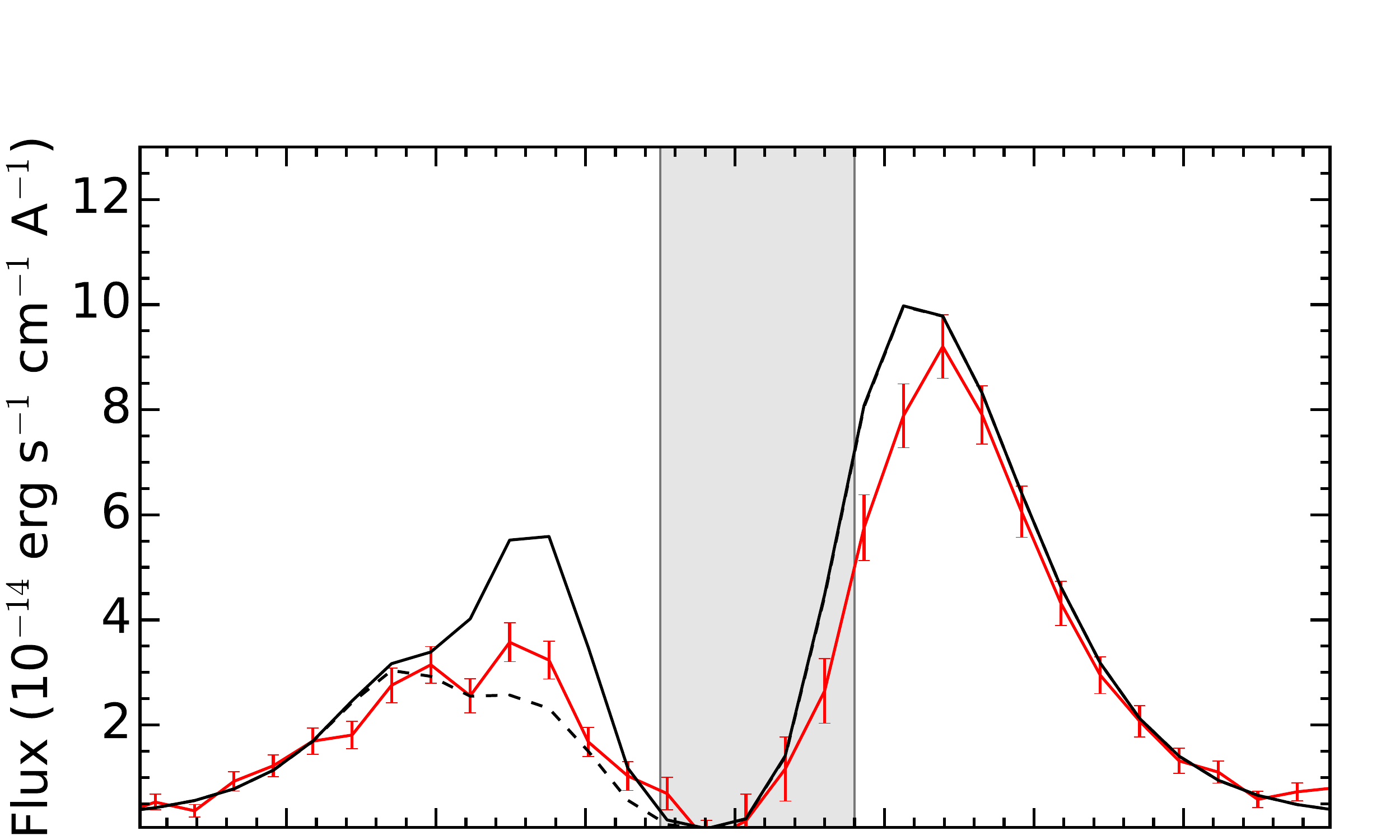}\\
\includegraphics[trim=0.3cm 0cm 1.2cm 0.1cm,clip=true,width=0.3545\textwidth]{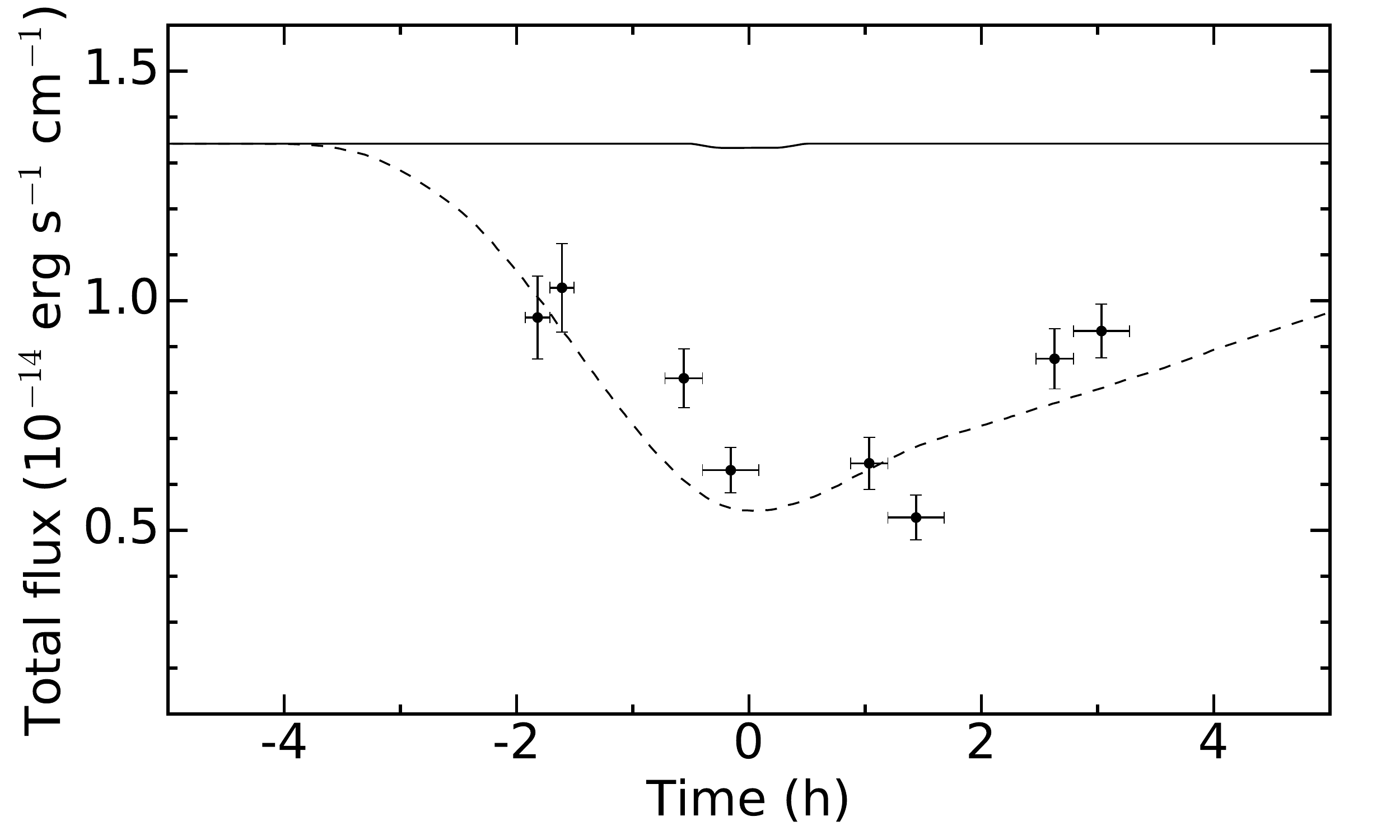}
\includegraphics[trim=2.9cm 0cm 1.2cm 0.4cm,clip=true,width=0.316\textwidth]{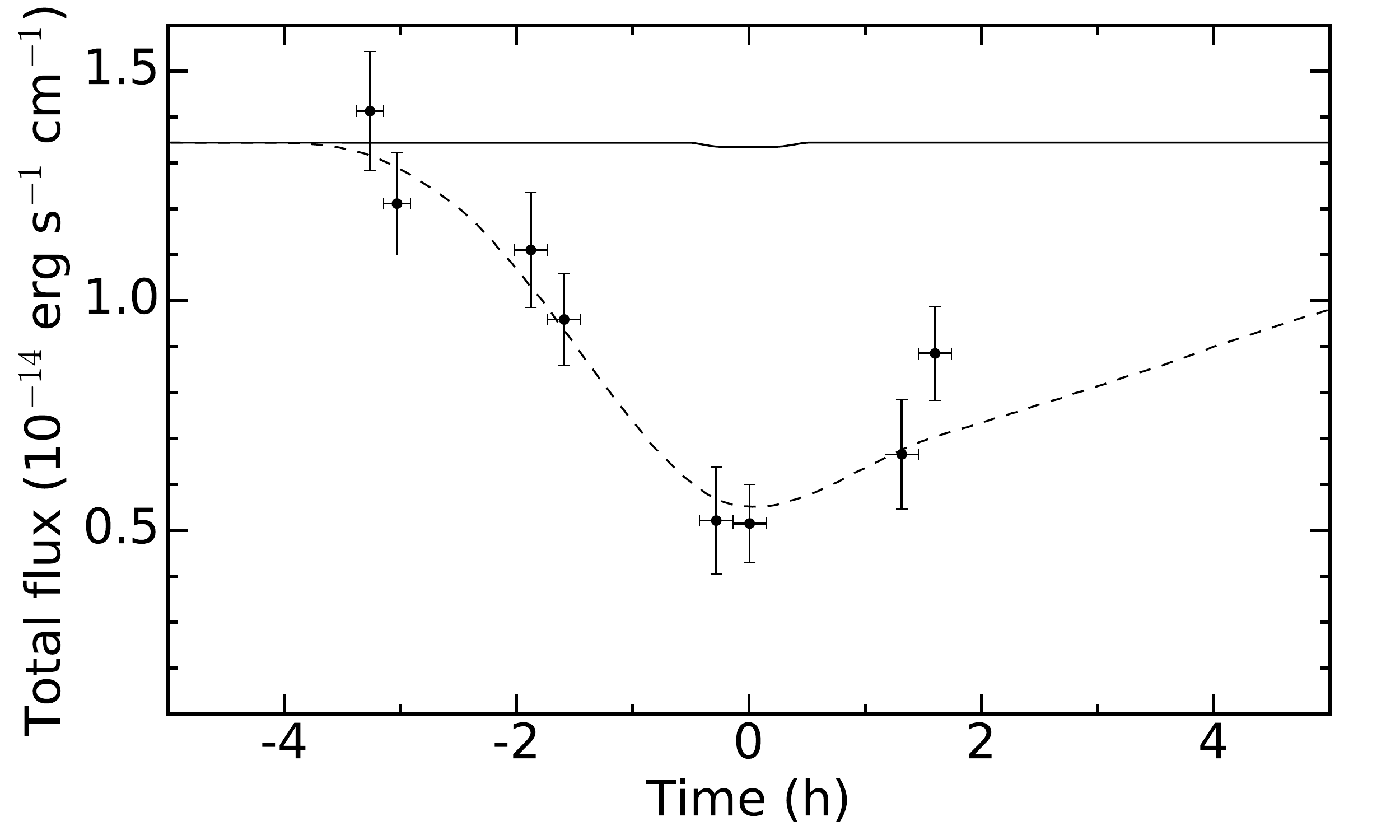}
\includegraphics[trim=2.9cm 0cm 1.2cm 0.4cm,clip=true,width=0.316\textwidth]{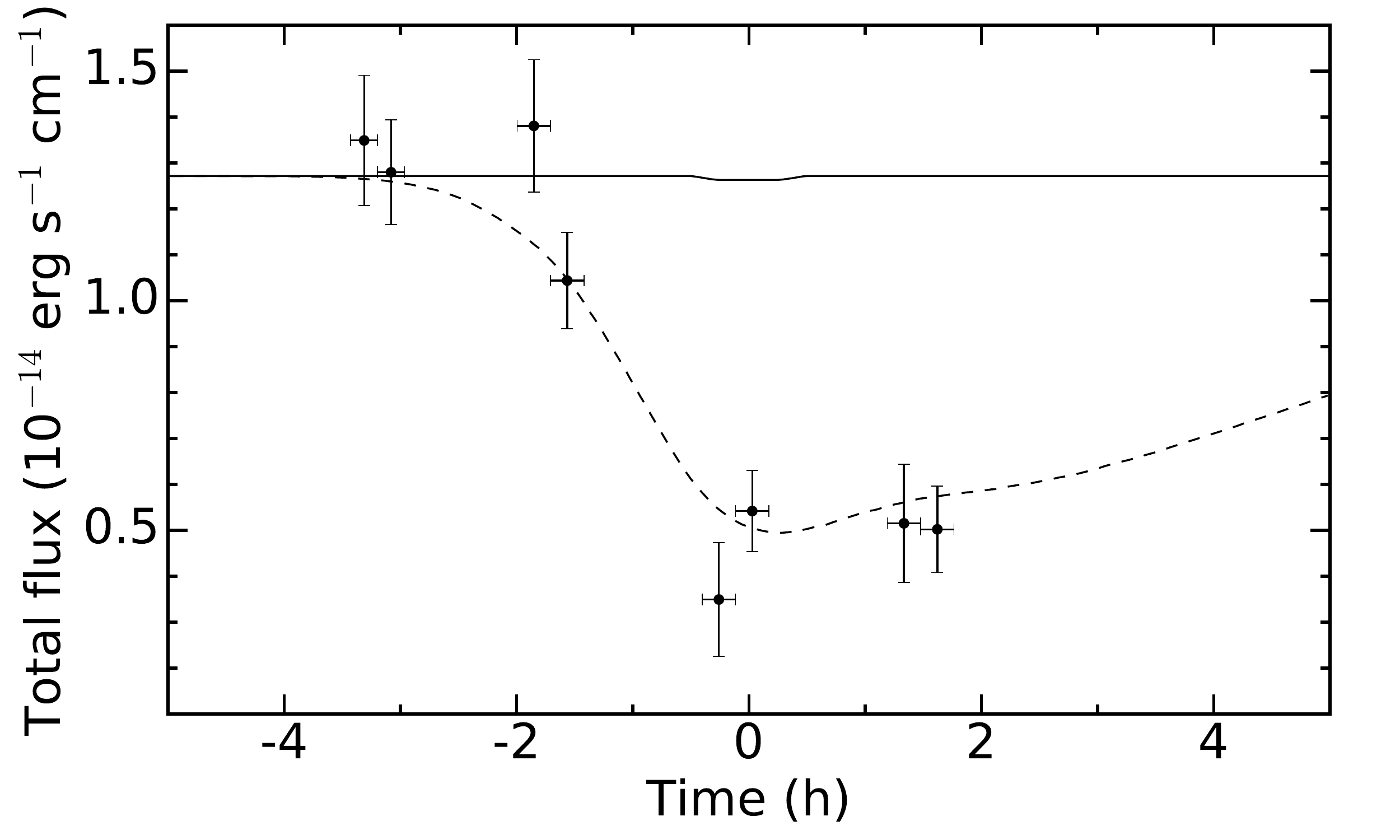}
\end{minipage}
\caption[]{GJ 436 spectra during Visits 1 (left column), 2 (central column), and 3 (right column). Spectra in the same row were measured at about the same phase of the planet orbital position. We show the spectra gathered over the full HST orbits. The shaded gray area corresponds to the range affected by ISM absorption and geocoronal emission, which was excluded from the fits. Black solid spectra are the reconstructed out-of-transit stellar line profiles. Dashed black spectra correspond to the best-fit theoretical spectra for Visits 2 (reported for comparison in Visit 1 epoch) and 3. Note that the transit of the exosphere has little effect on the red wing, where the best fits nearly overlap with the out-of-transit spectra. In the bottom part of the plot, light curves for each epoch show the observed and theoretical flux integrated between -120 and -40\,km\,s$^{-1}$. The solid black line shows the optical transit.}
\label{spectra}
\end{figure*}

The radiation pressure from the M dwarf GJ\,436 has a strong influence on the exosphere of its warm-Neptune companion and reproduces the velocity range of its absorption signature well (\citealt{Bourrier2015_GJ436}). However, radiative braking alone does not explain the variations of the absorption depth observed at the different phases of the transit well (Sect.~\ref{GJ436_obs}). While it allows the formation of a coma that is large enough to occult about half of the stellar disk at the center of the optical transit, this coma also extends too far ahead of the planet and produces a deeper and earlier ingress than observed (\citealt{Bourrier2015_GJ436}). Furthermore, each observation
epoch shows specific features that cannot be explained by a stable radiation pressure (see Fig.~\ref{spectra}). The flux in the blue wing of the Lyman-$\alpha$ line varies more smoothly over time during Visit 1, with a longer transit duration than other visits, but a lower absorption depth at the center of the transit. In contrast, Visit 2 shows sharper flux variations at the ingress and egress and a shorter, deeper transit. Finally, ingress starts the latest in Visit 3 with a dramatic increase in absorption depth, which surprisingly remains at about the same level during the transit and post-transit phases. This feature is at odds with the gradual decrease in absorption depth caused by stellar photoionization and the dilution of the gas subjected to radiative braking (\citealt{Bourrier2015_GJ436}).\\
An additional mechanism is therefore needed 1) to reduce the size of the coma ahead of the planet in all epochs and 2) to explain variations in the dynamics and the geometry of the exosphere between the different epochs. In this section, we show that by abrading the hydrogen cloud formed by planetary escape and by creating a secondary population of neutral hydrogen atoms, stellar wind interactions might be able to explain these two points. In Sect.~\ref{results} we compare observations and EVE simulations with a stable radiation pressure and a variable stellar wind and derive the corresponding properties of the stellar wind and planetary environment at the different epochs.

\subsubsection{Abrasion}
\label{sect:abrasion}

\begin{figure*}
\centering
\begin{minipage}[b]{\textwidth}
\includegraphics[trim=2cm 1cm 6.4cm 4cm,clip=true,width=0.531\columnwidth]{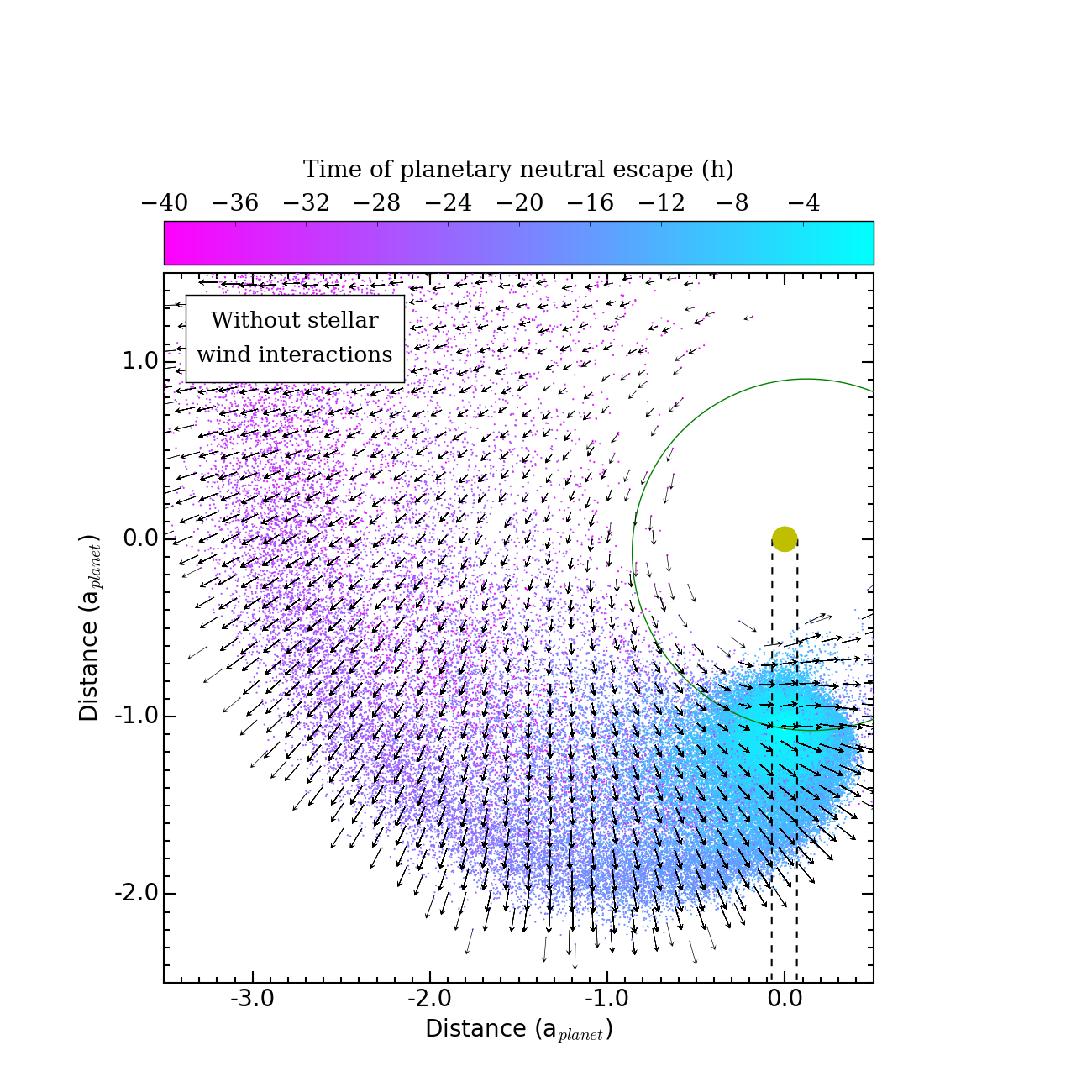}
\includegraphics[trim=4.83cm 1cm 6.4cm 4cm,clip=true,width=0.47\columnwidth]{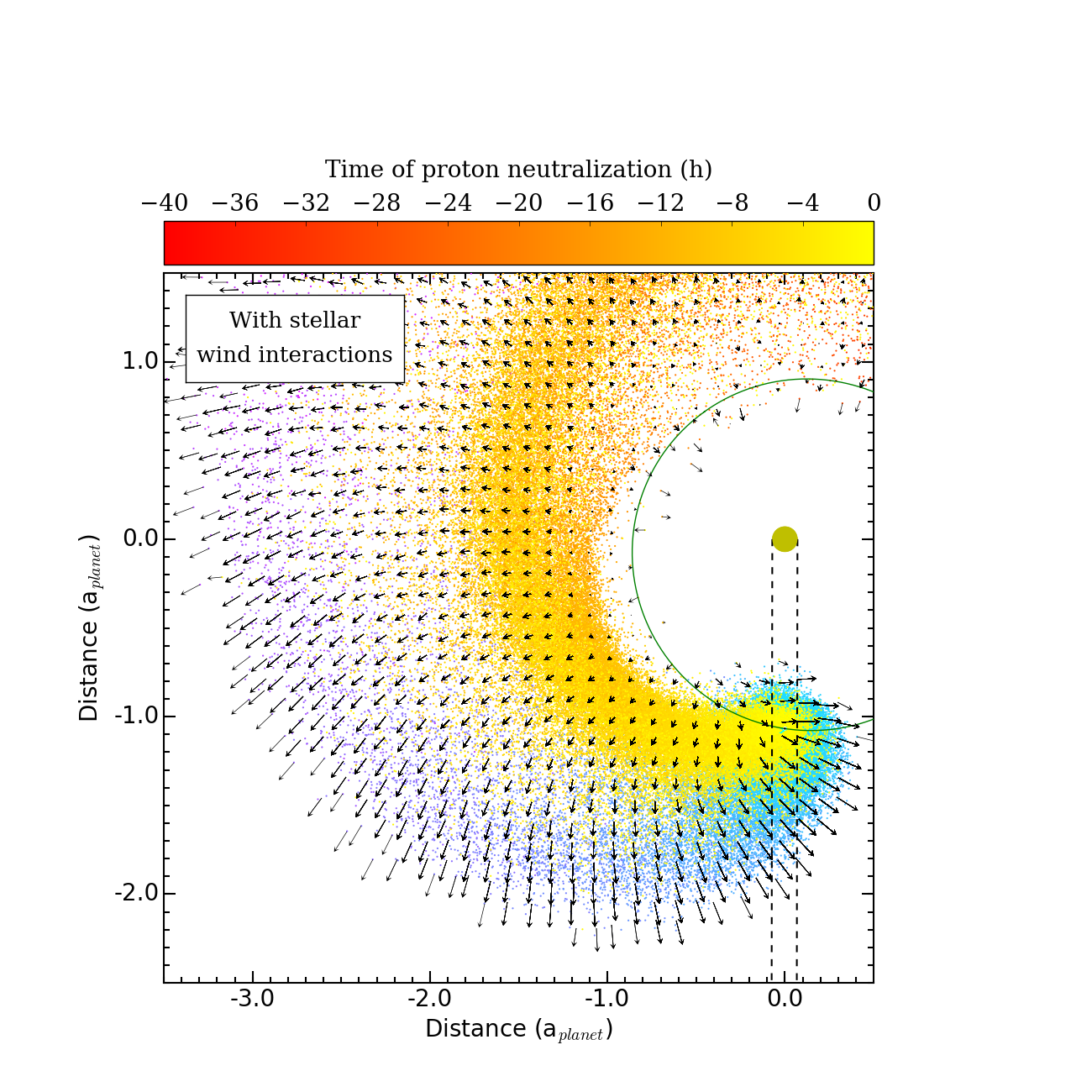}
\end{minipage}
\caption[]{Views of GJ\,436b exosphere (dots) within the orbital plane, seen from the top, at the time of the optical transit. The two panels correspond to two simulations performed with the same escape rate, planetary wind velocity, and photoionization rate, but without stellar wind interactions (\textit{left panel}) or including them (\textit{right panel}, corresponding to the best fit for Visit 2). The planet is represented by a small black disk, to scale, between the dashed black lines limiting the LOS toward the stellar disk. Arrows display the average velocity field of the neutral hydrogen atoms in the stellar rest frame, with particles colored as a function of the time they escaped the atmosphere (for planetary neutrals) or were created through charge exchange (for neutralized protons). With no stellar wind interactions the dynamics of the escaping gas is initially dominated by the planetary orbital velocity and later constrained by radiative braking. With charge exchange most protons are neutralized in the dense inner coma, abrading at the source planetary neutrals that would have fed the outer regions of the exosphere, but also creating a compact population of neutrals dominated by the radial bulk velocity of the stellar wind. It takes much more time for stellar gravity to eventually disperse this population of neutralized protons.}
\label{velfield_rad_stwind}
\end{figure*}

Through charge exchange, stellar wind protons ionize the neutral hydrogen exosphere arising from the planetary escape. Hereafter we refer to this phenomenon as abrasion to distinguish from the effect of stellar photoionization. The regions of the coma ahead of the planet and facing the star are strongly affected by abrasion, while self-shielding from the protons can protect the farthest regions opposite the star (Fig~\ref{cloud_schematics}). Nonetheless, our best-fit simulations (see Sect.~\ref{sec:cons_sp}) show that hydrogen densities are low enough around GJ\,436 b for the stellar wind to reach far within the core of the coma, interacting with many hydrogen atoms even before they can move away from the planet into the outer regions of the exosphere. Stellar wind abrasion thus also indirectly effects the cometary tail and the front of the coma, which are fueled by the expanding inner regions of the exosphere (see Figs.~\ref{cloud_schematics} and~\ref{velfield_rad_stwind}). This leads to a steeper decrease in absorption depth before and after the optical transit (Fig~\ref{dyn_time}).   \\

\subsubsection{Neutralized protons}
\label{sect:sec_pop}

\begin{figure*}
\begin{minipage}[b]{\textwidth}   
\hspace{0.006\textwidth}
\includegraphics[trim=0.15cm 0.12cm 0.49cm 0.cm,clip=true,width=0.326\textwidth]{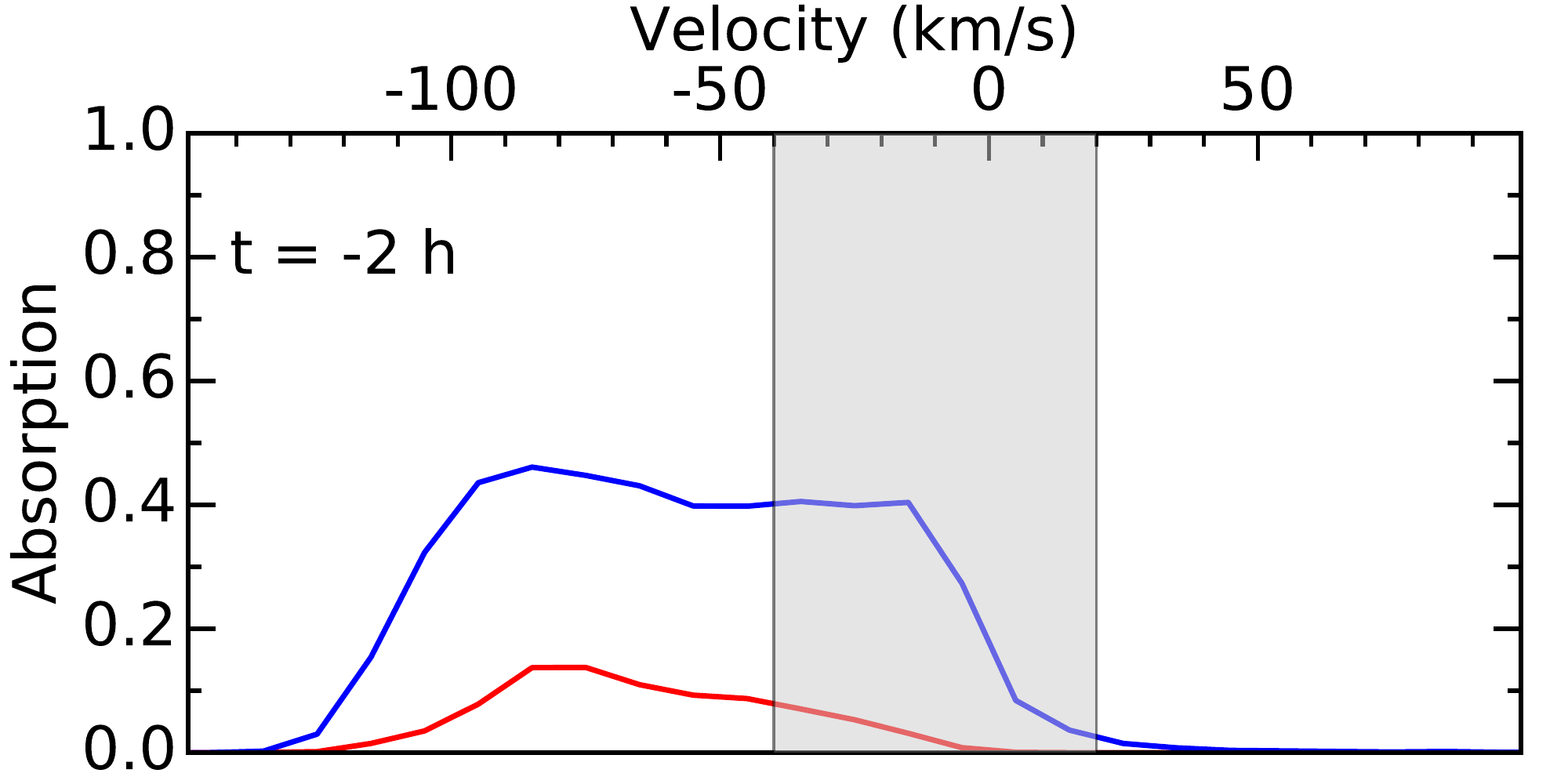}
\includegraphics[trim=2.3cm 0.1cm 0.46cm 0.cm,clip=true,width=0.29\textwidth]{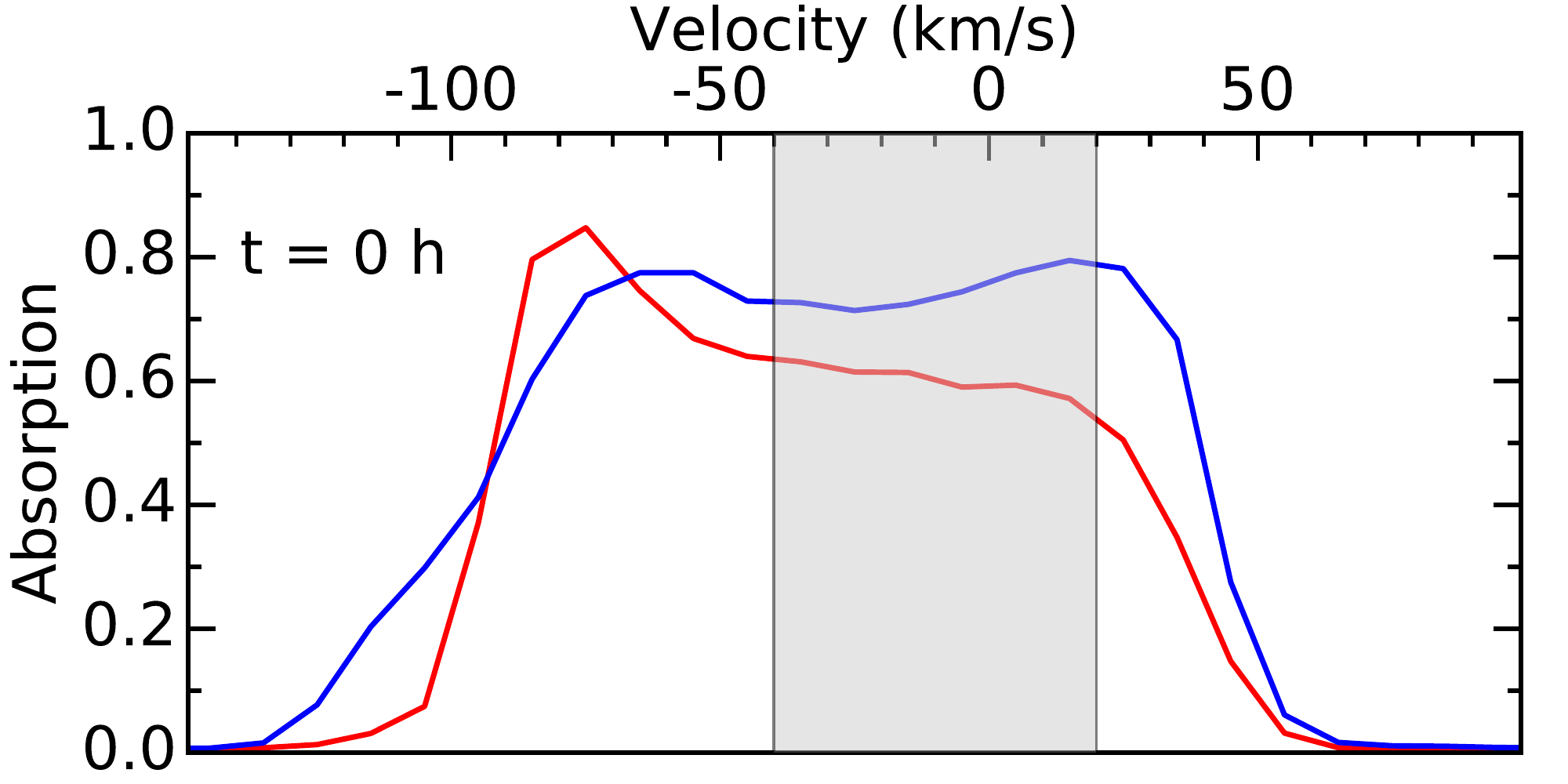}
\includegraphics[trim=2.3cm 0.1cm 0.49cm 0.cm,clip=true,width=0.289\textwidth]{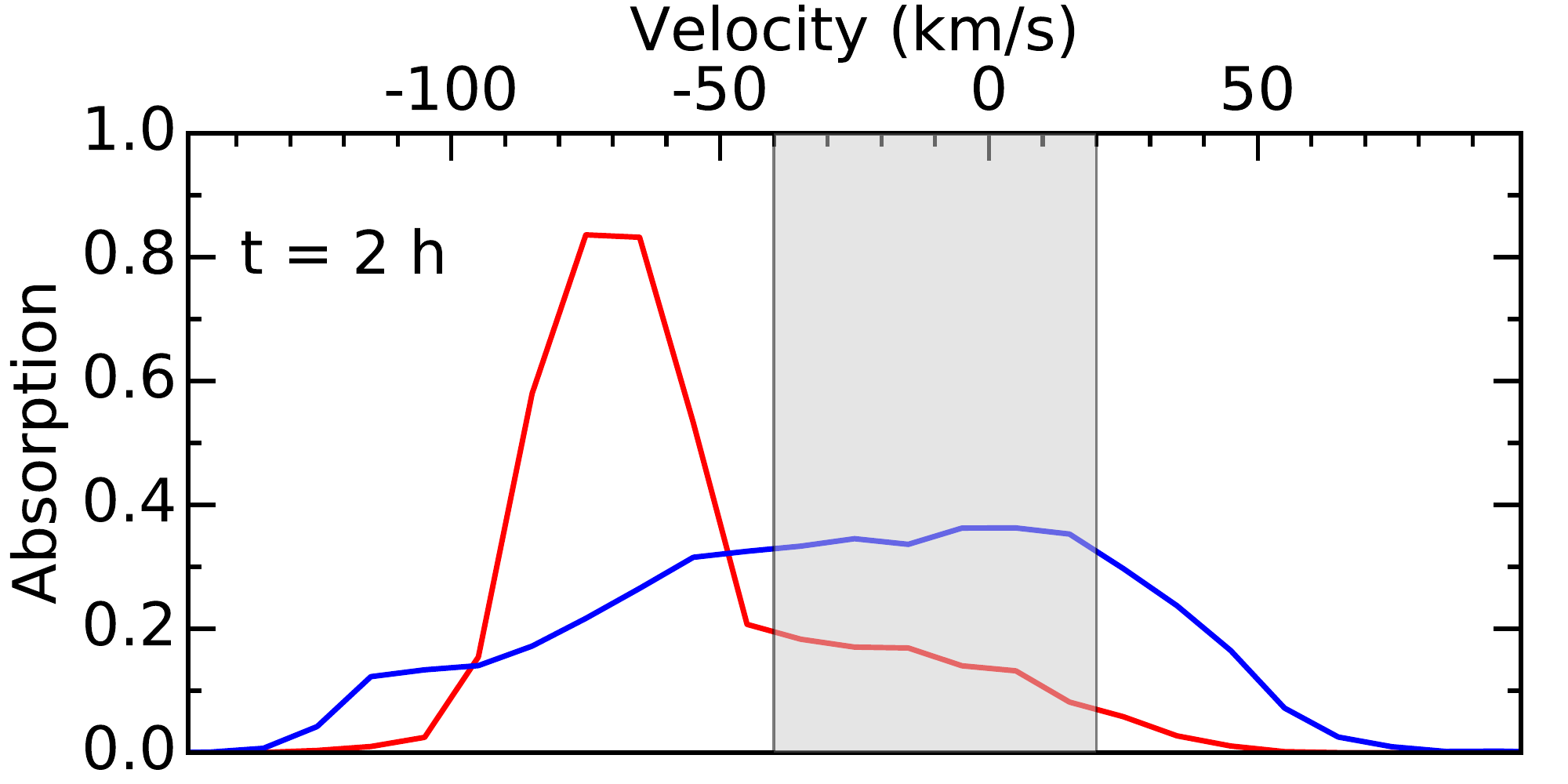}\\
\includegraphics[trim=1.2cm 1.2cm 6.4cm 7.8cm,clip=true,width=0.339\textwidth]{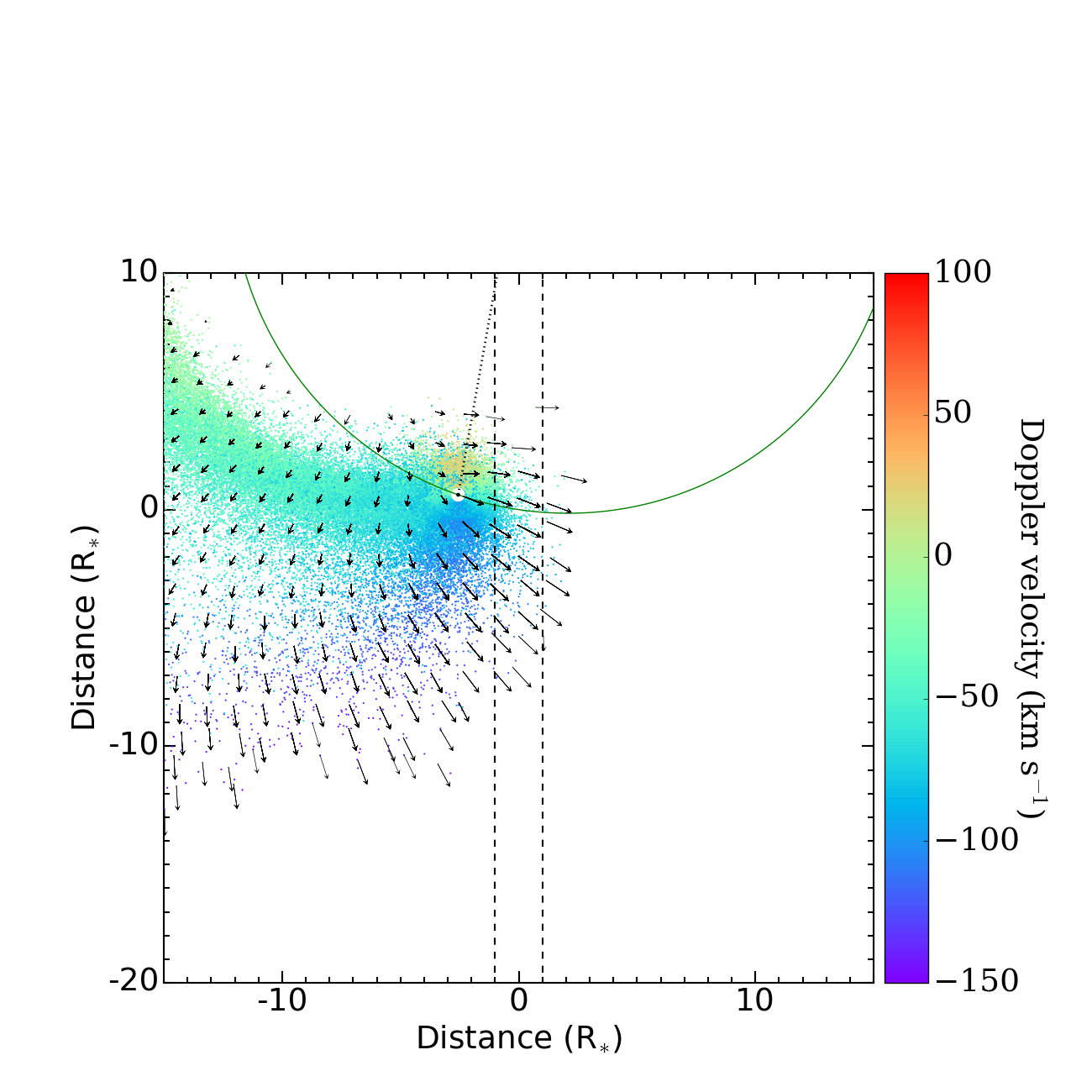}
\includegraphics[trim=4.85cm 1.2cm 6.4cm 7.85cm,clip=true,width=0.2905\textwidth]{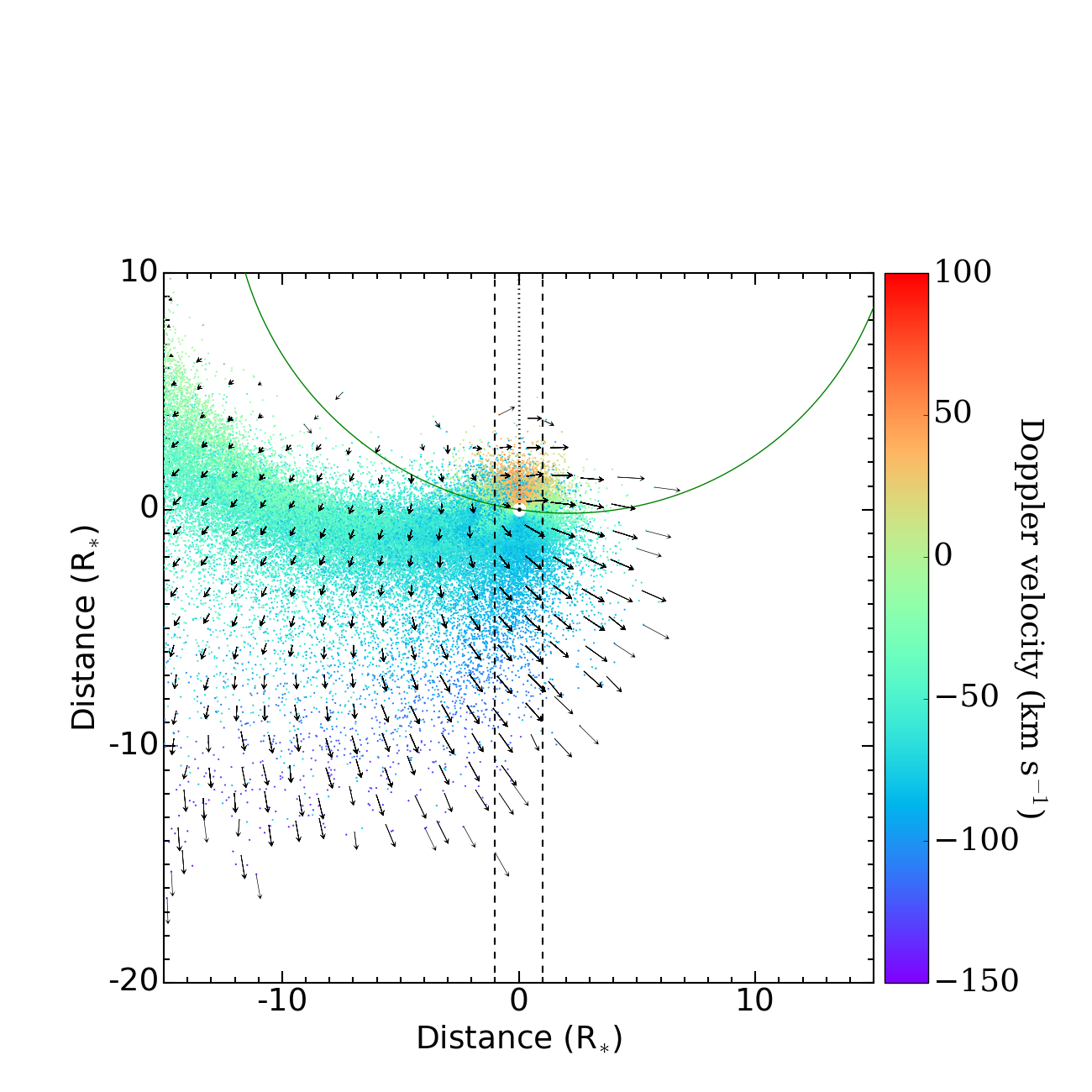}
\includegraphics[trim=4.85cm 1.2cm 1.4cm 7.85cm,clip=true,width=0.357\textwidth]{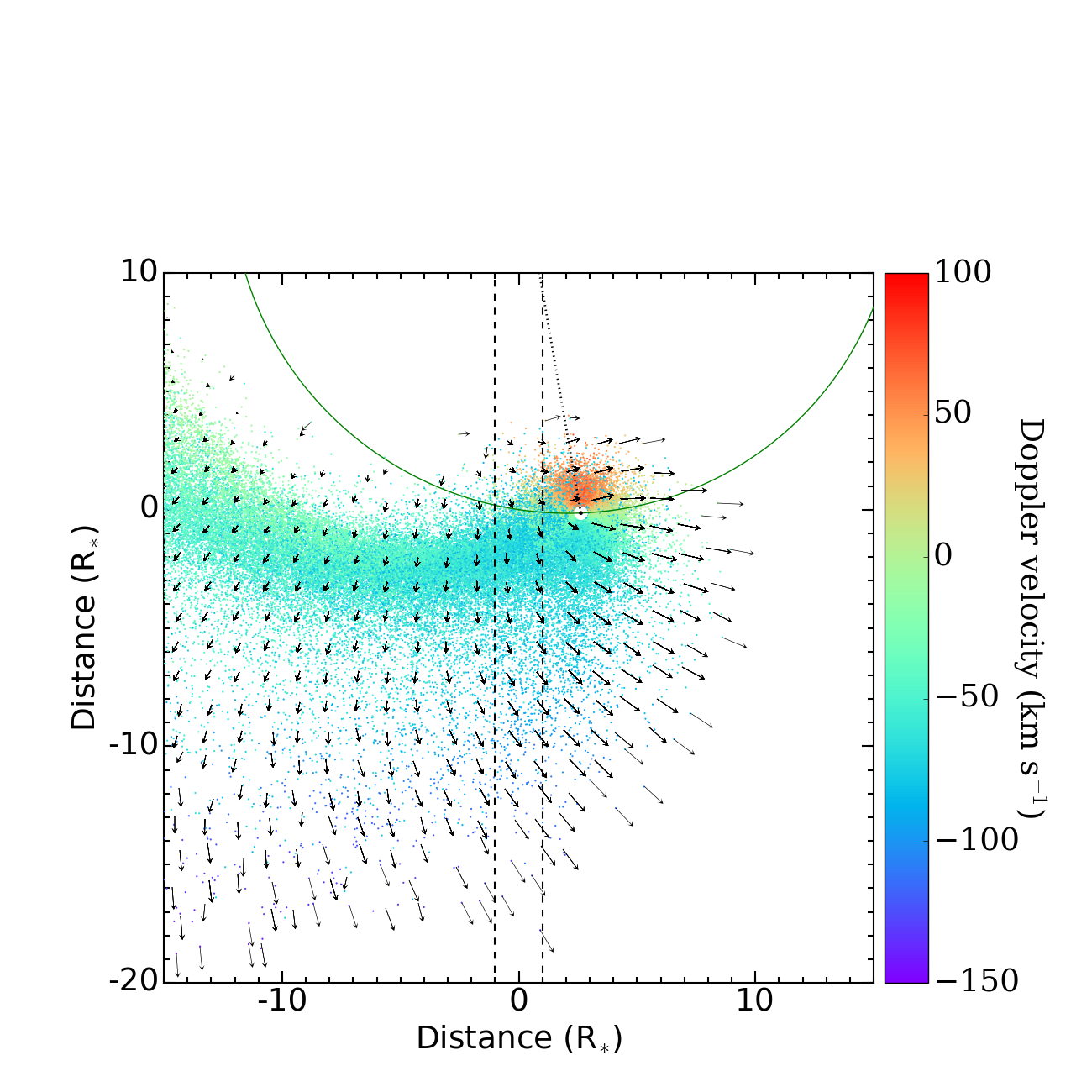}\\
\end{minipage}
\caption[]{Views of GJ\,436b exosphere (\textit{lower panels}) and its theoretical absorption profile (\textit{upper panels}). We caution that absorption in the shaded range ([-40 ; 20] km\,s$^{-1}$) cannot be observed from Earth and that the STIS LSF would spread the absorption signal from the exosphere over more data points. Time increases from left to right and is indicated in each plot. The dashed black lines limit the LOS toward the stellar disk, while the dotted line indicates the star-planet axis. Black arrows show the velocity field of the gas in the stellar rest frame, with H$^{0}$ atoms in the orbital plane colored as a function of their LOS-projected velocity. Red lines in the \textit{upper panels} show the absorption profiles from the exosphere simulated with Visit 3 best-fit parameters. The population of neutralized protons arising from the inner coma can be identified by its radial velocities and stable high-velocity absorption peak, while planetary neutrals remain present at the front of the coma and in the outer regions of the exosphere opposite the star with dynamics dominated by radiative braking. The blue line profiles in the \textit{upper panels} correspond to the same simulation with no stellar wind, illustrating how charge exchange reduces the size of the exosphere before and after the optical transit but only weakly affects the absorption profile at the center of the transit. It also shows that neutralized protons produce a stable high-velocity absorption peak.}
\label{dyn_time}
\end{figure*}

In addition to abrading neutral hydrogen atoms in the planetary exosphere, another effect of charge-exchange interactions is to neutralize protons in the stellar wind. This leads to the formation of a secondary population of neutral hydrogen in the exosphere (hereafter, neutralized protons), which differs from the population of neutral hydrogen atoms that escaped from the upper atmosphere of the planet (hereafter, planetary neutrals). This population of neutralized protons has two important characteristics. First, it arises primarily from the inner coma of GJ\,436b where most protons are neutralized because of the higher density of planetary neutrals close to the planet (Fig.~\ref{velfield_rad_stwind}, right panel). Then, planetary neutrals escaping the atmosphere initially have a low radial velocity that naturally increases as they move away from the planet orbit\footnote{The velocity of escaping atoms is dominated by the near-tangential orbital velocity of the planet in the stellar rest frame; because of radiative braking, these atoms decelerate with respect to the planet and the radial projection of their velocity thus slowly increases; see \citet{Bourrier2015_GJ436} for more details.}. In contrast, neutralized protons continue to move with the velocity distribution of the stellar wind, which is dominated by a high radial bulk velocity. Consequently, while it takes several hours for planetary neutrals to move into the outer regions of the exosphere, neutralized protons swiftly move away from the planet and can go farther before they are photoionized. Because of these two features, the population of neutralized protons is shaped into a compact cloud originating from the core of the coma and extending into a long comet-like tail that moves with the persistent velocity distribution of the stellar wind (Fig.~\ref{velfield_rad_stwind}, right panel). This tail spreads more slowly than the planetary one, both dynamically and geometrically (Fig.~\ref{velfield_rad_stwind}, see left and right panel). For several hours after the optical transit the population of neutralized protons thus produces a secondary absorption profile with a fairly stable spectral range and slowly decreasing depth (Fig.~\ref{dyn_time}). Eventually, this profile spreads and shifts toward more positive radial velocities as stellar gravity slows down atoms in the tail and creates a strong velocity gradient between its outer C-shaped region and the inner C-shaped region closer to the star (Fig.~\ref{velfield_rad_stwind}, right panel). We caution that for the absorption signature of neutralized protons to become directly observable, their bulk velocity must be lower than about 200\,km\,s$^{-1}$. With higher absolute velocity there is not enough flux in the Lyman-$\alpha$ line of GJ\,436 (see Fig.~\ref{spectra}), and only the abrading effect of the wind on the exosphere can be detected.

\subsection{Influence of the physical parameters}
\label{sec:influence}

We describe here the first-order influence of the escape rate and velocity of the planetary outflow and of the stellar photoionization rate on the structure of the exosphere and its transmission spectrum when it is subjected to stellar wind interactions. We describe the role of each parameter of the modeled stellar wind in more detail (previous descriptions in the literature can be found in \citealt{Holmstrom2008}; \citealt{Ekenback2010}; \citealt{Bourrier_lecav2013} , \citealt{Kislyakova2014}).

\begin{itemize}
\item \textit{Planetary wind velocity $v^{\mathrm{p}}_{\mathrm{wind}}$}: This parameter corresponds to the initial upward bulk velocity of the atoms escaping the upper atmosphere and influences the spatial and spectral dispersion of the gas in the exosphere. With higher escape velocities, the wider distribution of Doppler velocities for the planetary neutrals spreads their absorption signature over a broader spectral range, but the velocity distribution of the neutralized protons remains that of the stellar wind (Sect.~\ref{st_wind_model}) and their absorption profile covers the same spectral range. Nonetheless, the effective variation in absorption depth with time and wavelength is complex and depends on the column densities and stellar disk area that is occulted by neutral hydrogen coming from both populations. For example, an increase in $v^{\mathrm{p}}_{\mathrm{wind}}$ expands the coma of planetary neutrals and increases its dilution. Stellar wind protons are thus neutralized within a larger volume, but with lower local densities.\\

\item \textit{Photoionization rate $\Gamma_{\mathrm{ion}}$:} Photoionization leads to a general decrease in density and absorption from all regions of the exosphere, albeit with differences caused by self-shielding and the influence of stellar wind interactions. The structure of the exosphere is less affected by the photoionization of neutralized protons than planetary neutrals. The latter move through the exosphere on longer timescales, making their population more vulnerable to photoionization, which additionally removes the neutral planetary material required for charge exchange with the stellar wind (Sect.~\ref{sect:sec_pop}). \\

\item \textit{Escape rate $\dot{M}_{\mathrm{H^{0}}}$:} Higher values for the escape rate increase the densities of planetary neutrals and neutralized protons in the exosphere, which deepens the absorption profiles of both populations. Different regions of the exosphere have different velocity distributions and contribute to the absorption in different spectral ranges (Sect.~\ref{sect:sec_pop}). However, self-shielding remains low enough for GJ\,436b that variations in the escape rate propagate roughly uniformly throughout the whole exosphere and do not affect the velocity distribution of the gas and the spectral range of its absorption profile. We caution that variations related to $\dot{M}_{\mathrm{H^{0}}}$ would be different for saturation regimes of the stellar wind, in which either all escaping planetary neutrals would interact with the incoming protons (as for HD\,189733b, \citealt{Bourrier_lecav2013}), or all protons in the stellar wind would be neutralized by the planetary exosphere. Simulations show, however, that GJ\,436 b is not observed in such regimes (Sect.~\ref{sec:cons_sp}).\\

\item \textit{Stellar proton density $n^{\mathrm{st}}_{\mathrm{wind}}$:} This parameter influences the number of planetary neutrals abraded by charge exchange and consequently the number of neutralized stellar wind protons. A higher stellar wind density increases the abrasion of the exosphere, which reduces the size of the exosphere envelop and the absorption depth before and after the optical transit (Sect.~\ref{sect:abrasion}). This can be partly compensated for by the neutralized protons that contribute to the observed Lyman-$\alpha$ absorption in the spectral range that corresponds to the stellar wind velocity distribution. This additional absorption is mainly visible during and after the optical transit because protons neutralized ahead of the planet move away into the tail much more quickly than the planetary neutrals (Sect.~\ref{sect:sec_pop}).\\

\item \textit{Stellar proton kinetic temperature $T^{\mathrm{st}}_{\mathrm{wind}}$:} The kinetic temperature of the stellar wind controls the spread of the protons velocity distribution and thus the breadth of the neutralized protons absorption profile population. A higher temperature decreases the absorption depth in the core of the profile in favor of its wings. The overall effect on the exosphere is stronger after the optical transit because a higher dispersion of the protons neutralized in the coma feeds back the spatial and spectral dispersion of the gas in the cometary tail. The value of $T^{\mathrm{st}}_{\mathrm{wind}}$ has very little influence on the abrasion of the exosphere because the cross section for charge exchange varies slowly with the relative velocities at play between protons and planetary neutrals (Sect.~\ref{st_wind_model}), and their probability of interaction therefore does not depend strongly on the thermal dispersion of the proton velocity.   \\

\item \textit{Stellar proton velocity $V^{\mathrm{st}}_{\mathrm{bulk-wind}}$:} The velocity distribution of the stellar wind is dominated by its radial bulk velocity. Therefore the projected velocity of the neutralized protons is close to their absolute velocity in the stellar rest frame when they are transiting the star, and $V^{\mathrm{st}}_{\mathrm{bulk-wind}}$ determines the center of their absorption profile in the spectra. By contrast, the velocity distribution of the planetary neutrals is strongly dependent on the planet orbital velocity, and their absorption profile is closer to the core of the Lyman-$\alpha$ line (\citealt{Bourrier2015_GJ436}). Regarding the abrasion of the exosphere, variations in the bulk velocity affect both the relative velocities between protons and planetary neutrals and their interaction cross-section (Sect.~\ref{st_wind_model}). An increase in $V^{\mathrm{st}}_{\mathrm{bulk-wind}}$ tends to slightly increase the probability for a planetary neutral to undergo charge exchange. \\

\end{itemize}


\section{Measuring the properties of GJ\,436b environment}
\label{results}

\subsection{Stellar XEUV emission and photoionization rate}
\label{st_emission}

We searched for prior constraints to place on the EVE parameters used to interpret the Lyman-$\alpha$ line observations of GJ\,436 b. It is not possible to use the planetary mass-loss properties derived by \citet{Kulow2014}, which are biased by their interpretation of Visit 1 (Sect.~\ref{GJ436_obs}). \citet{Ehrenreich2015} did not account for possible temporal variability between each visit and made a strong assumption on the presence of a sharp density transition in the outer regions of the exosphere facing the star. \citet{Bourrier2015_GJ436} obtained rough estimates for the planetary outflow velocity, but with the assumption that radiation pressure alone acts on the exosphere. Regarding the host star, no measurements of the stellar wind are available. However, \citet{Bourrier2015_GJ436} reconstructed the intrinsic stellar Lyman-$\alpha$ line for Visits 2 and 3, and this can be used to estimate the EUV spectrum of GJ\,436 and the resulting photoionization rate per neutral hydrogen atom at 1\,au from the star:   

\begin{equation}
\widetilde{\Gamma}_{\mathrm{ion}}=\int^{911.8\,\AA} \! \frac{F_{EUV}(\lambda)\,\sigma_{ion}(\lambda)}{hc} \, \lambda\, \mathrm{d}\lambda, 
\end{equation} 
with $\widetilde{\Gamma}_{\mathrm{ion}}$ in s$^{-1}$, $F_{EUV}(\lambda)$ the stellar flux at 1\,au (in erg\,s$^{-1}$\,cm$^{-2}$\,\AA$^{-1}$) and $\sigma_{ion}$ the cross section for photoionization (in cm$^{2}$). The integration was performed up to the ionization threshold at 911.8\,\AA. The cross section for hydrogen photoionization is wavelength dependent, and we used the expression from \citet{Verner1996} and \citet{Bzowski2013}:
\begin{equation}
\sigma_{\mathrm{ion}}=6.538\times10^{-32}\,\left( \frac{29.62}{\sqrt{\lambda}} +1 \right)^{-2.963}\,(\lambda-28846.9)^{2}\,\lambda^{2.0185},
\end{equation}
with $\sigma_{\mathrm{ion}}$ in cm$^2$ and wavelengths in \AA. The X-ray emission of GJ\,436 from 5 to 100\,\AA\, was measured with Chandra at the epoch of Visit 2 (\citealt{Ehrenreich2015}). The EUV emission from 100 to 912\,\AA\, is mostly absorbed by the ISM and must be estimated through indirect methods. We used the scaling relations of \citet{Linsky2014}\footnote{We caution that the coefficients for F5-M5\,V stars in \citet{Linsky2014}, Table 5, must be used with the logarithm of the stellar Lyman-$\alpha$ flux at 1\,au, multiplied by the scale factor $(R_\mathrm{\sun}/R_{*})^2$.}, which are based on the integrated intrinsic Lyman-$\alpha$ flux at 1\,au from GJ\,436 that we measured to be 0.90$\stackrel{+0.20}{_{-0.14}}$\,erg\,s$^{-1}$\,cm$^{-2}$ for Visit 2 and 0.88$\stackrel{+0.30}{_{-0.21}}$\,erg\,s$^{-1}$\,cm$^{-2}$ for Visit 3. The uncertainties on these values were obtained by varying the peak flux of the Lyman-$\alpha$ line constrained by \citet{Bourrier2015_GJ436} within its 1$\sigma$ error bars. We note that \citet{France2013} performed a similar estimation using the intrinsic Lyman-$\alpha$ line of GJ\,436 observed in 2010 (\citealt{Ehrenreich2011}). With a maximum uncertainty of 30\%, the total flux they obtained ranges from 1.1 to 2.0\,erg\,s$^{-1}$\,cm$^{-2}$ at 1\,au from the star. This is marginally higher ($\sim$1.2$\sigma$) than our estimates, possibly because of long-term variations of the Lyman-$\alpha$ line flux between the 2 - 3.5 years that separate this earlier observation from Visits 2 and 3, or because of short-term temporal variability through impulsive flares in chromospheric and transition region emission lines of the host star at this epoch (\citealt{France2013}). \\

Our results for the stellar flux and luminosity in Visits 2 and 3 are shown in Table~\ref{XEUV_flux} for complementary wavelength bands of the XEUV domain. Uncertainties on the integrated Lyman-$\alpha$ flux were propagated on the flux in each domain to estimate 1$\sigma$ error bars on $\widetilde{\Gamma}_{\mathrm{ion}}$. We neglected uncertainties on the X-ray emission because it is about four times lower than the EUV emission and $\sigma_{\mathrm{ion}}$ steeply decreases with $\lambda\approxinf$100\AA, making the contribution of the X-ray flux to the photoionization rate negligible. We found photoionization rates at 1\,au from the star of 1.9$\stackrel{+0.7}{_{-0.4}}\times$10$^{-8}$\,s$^{-1}$ for Visit 2 and 1.8$\stackrel{+1.1}{_{-0.6}}\times$10$^{-8}$\,s$^{-1}$ for Visit 3. This corresponds to 2.3$\stackrel{+0.8}{_{-0.5}}\times$10$^{-5}$\,s$^{-1}$ and 2.2$\stackrel{+1.3}{_{-0.7}}\times$10$^{-5}$\,s$^{-1}$ at the distance of the semi-major axis. The values for both visits are remarkably similar, as expected from the very stable Lyman-$\alpha$ line profiles between the two epochs (\citealt{Ehrenreich2015}; \citealt{Bourrier2015_GJ436}). Because of this stability and the short six-month interval between Visits 1 and 2, we assumed these photoionization rates to be a good estimate for Visit 1.\\

\begin{table*}
\centering
\caption{X-EUV emission of GJ\,436b}
\begin{tabular}{lcccccc}
\hline
\hline
\noalign{\smallskip}    
Wavelengths         & \multicolumn{2}{c}{Stellar flux at 1\,au} & \multicolumn{2}{c}{Stellar flux at the semi-major axis}  & \multicolumn{2}{c}{Stellar luminosity}  \\      
\noalign{\smallskip}
(\AA)               & \multicolumn{2}{c}{(erg\,s$^{-1}$\,cm$^{-2}$)} & \multicolumn{2}{c}{(erg\,s$^{-1}$\,cm$^{-2}$)}  & \multicolumn{2}{c}{(10$^{26}$erg\,s$^{-1}$)}         \\      
\noalign{\smallskip}
                     &  Visit 2                 & Visit 3     &  Visit 2             & Visit 3       &    Visit 2            & Visit 3      \\ 
\noalign{\smallskip}
\hline
5 - 100$^{\dagger}$  & 0.205                    &       -         & 249.0   & -          & 5.77                          &  -  \\ 
100 - 200                       & 0.290                         & 0.283     & 351.4   & 344.0                     &  8.14                         & 7.97\\
200 - 300                       & 0.254                         & 0.249     & 308.2   & 301.7                     &  7.14                         & 6.99 \\ 
300 - 400                       & 0.224                         & 0.219     & 272.2   & 266.4                     & 6.30                          & 6.17 \\
400 - 500                       & 0.007                         & 0.007     & 8.2   & 8.0                 &  0.19                         & 0.19 \\ 
500 - 600                       & 0.017                         & 0.017     & 21.0   &     20.3           & 0.49                          & 0.47 \\
600 - 700                       & 0.016                         & 0.015     & 19.0   & 18.5                       & 0.44                          & 0.43 \\ 
700 - 800                       & 0.025                         & 0.025     & 31.0   &     30.0           & 0.72                          & 0.69  \\
800 - 912                       & 0.045                         & 0.043     & 54.8   &     52.8           & 1.27                          & 1.22 \\ 
912 - 1170                      & 0.085                         & 0.083     & 102.8   & 100.6                     & 2.38                  & 2.33  \\
Lyman-$\alpha$          & 0.897                                 & 0.878     & 1088.5   & 1065.5                   & 25.22                 & 24.68 \\
\noalign{\smallskip}
\hline
\hline
\multicolumn{7}{l}{$\dagger$: X-ray emission from Chandra measurements (\citealt{Ehrenreich2015})} \\
\end{tabular}
\label{XEUV_flux}
\end{table*}


\subsection{Constraints from the observed spectra}
\label{sec:cons_sp}

Using the settings described in Sect.~\ref{model}, we compared spectra from EVE simulations with Lyman-$\alpha$ observations to measure the properties of GJ\,436 b environment that best explain the observations. The spectra in each visit were fit independently to strengthen the reliability of the estimations and test the scenario with different data sets. Because of the runtime for a single simulation (from $\sim$5\,hours to several days), it is not practical to explore the parameter space using MCMC algorithms. Instead, we computed the $\chi^2$ of the fits on a grid scanning all possible values for the six model parameters (see Sect.~\ref{model}): the escape rate of neutral hydrogen $\dot{M}_{\mathrm{H^{0}}}$, the planetary outflow velocity $v^{\mathrm{p}}_{\mathrm{wind}}$, the photoionization rate $\Gamma_{\mathrm{ion}}$, and the stellar wind properties (bulk velocity $V^{\mathrm{st}}_{\mathrm{bulk-wind}}$, kinetic dispersion $v^{\mathrm{st}}_{\mathrm{therm-wind}}$, and density $n^{\mathrm{st}}_{\mathrm{wind}}$ of the protons distribution). When the absolute minimum $\chi^2$ and corresponding best values for the parameters were obtained, we calculated their error bars from an analysis of $\chi^2$ variations. A given parameter was pegged at various trial values, and for each trial value we searched for the minimum $\chi^2$ with the five other parameters that were allowed to vary freely. The 1$\sigma$ error bar for the pegged parameter was obtained when its value yields a $\chi^2$ increase of 1 from the absolute minimum (see, e.g., \citealt{hebrard2002}). We used the independent estimates of the photoionization rate $\widetilde{\Gamma}_{\mathrm{ion}} \pm \sigma_{\widetilde{\Gamma}_{\mathrm{ion}}}$ derived in Sect.~\ref{st_emission} as constraints on the model value $\Gamma_{\mathrm{ion}}$, adding to the $\chi^2$ the term $ ( ( \Gamma_{\mathrm{ion}} - \widetilde{\Gamma}_{\mathrm{ion}} )/\sigma_{\widetilde{\Gamma}_{\mathrm{ion}}} )^2 $.\\
In a first scenario (Sect.~\ref{sect:lowvel_results}), we explored moderate values for the stellar wind bulk velocities in the range of the Lyman-$\alpha$ line Doppler width ($\approxinf$200\,km\,s$^{-1}$) to allow for both abrasion and proton neutralization in the exosphere. Abrasion alone from a high-velocity stellar wind ($\approxsup$200\,km\,s$^{-1}$) is discussed as a second scenario in Sect.~\ref{high_wind}. 

\subsubsection{Stellar wind abrasion and proton neutralization}
\label{sect:lowvel_results}

Best-fit values for Visits 2 and 3 are given in Table~\ref{results_fits}, along with their 1$\sigma$ uncertainties. They provide a good fit to the data, with $\chi^2$ of 203 and 182 for 200 and 198 degrees of freedom (dof), respectively. Observations in both epochs require significant interactions between the exosphere and the stellar wind, with proton densities in the order of 10$^{3}$\,cm$^{-3}$ at the location of the planet. These interactions abrade the coma and produce a transit ingress starting $\sim$3\,h before the optical transit center, consistent with the observations (Fig.~\ref{velfield_rad_stwind}). Without these interactions, the ingress would start too early, about 5\,h before the optical transit center (\citealt{Bourrier2015_GJ436}). The proton density is higher in Visit 3 than in Visit 2, which is consistent with a stronger abrasion causing the sharper ingress observed during Visit 3 (Sect.~\ref{sect:abrasion}) and with a more abundant population of neutralized protons in the exosphere producing a stable absorption signature at later orbital phases (Sect.~\ref{sect:sec_pop}). Except for variations in the proton density and a marginally lower atmospheric escape velocity in Visit~2, the planetary outflow and stellar properties inferred for the GJ\,436 system are very similar in both epochs, with all parameters consistent at the 1$\sigma$ level. The consistency between these parameters, derived from independent datasets obtained at two different epochs, gives credence to the present scenario and shows the stability of the planetary mass loss over time. We note that removing the prior constraint on the photoionization rate has little influence on the quality of the fits over a broad range of values because it leads to changes in the local density of planetary neutrals that can be compensated for by variations in the escape rate and/or in the proton density.\\

We found that we were unable to perform a $\chi^2$ analysis for Visit 1, with no clear minimum and spurious $\chi^2$ variations that did not allow us to constrain the parameter values at this epoch. Using the lines reconstructed for Visits 2 and 3 as reference for Visit 1 may have biased the evaluation of radiation pressure and the calculation of the theoretical transmission spectra for this epoch. We also investigated whether these problems might have been caused by the flux variations in the red wing of the Lyman-$\alpha$ line (Fig.~\ref{spectra}), since absorption at positive velocities cannot not be explained by stellar wind interactions. But limiting the fit to the blue wing of the line did not improve the $\chi^2$ analysis. It is possible that these variations in the shape of the line are caused by active Lyman-$\alpha$ regions at the surface of the stellar disk (\citealt{Llama2016}), although this is made unlikely by the very localized spectral ranges of these variations and the stability of the Lyman-$\alpha$ line over time (Fig.\ref{spectra}). Finally, we compared the best
fits to Visits 2 and 3 (Table~\ref{results_fits}) with Visit 1 observations and obtained $\chi^2$ values of 204 and 212 for 104 points in the blue wing (445 and 447 for 208 points in both wings). Although Visit 1 shares similarities with the other epochs (see \citealt{Ehrenreich2015}), this hints at more drastic differences in the physical conditions of GJ\,436b exosphere during the first epoch. \\


\begin{table*}
\caption{Best-fit parameters and 1$\sigma$ uncertainties derived from the fits to the Lyman-$\alpha$ line observations of GJ\,436 b in Visits 2 and 3.}
\begin{tabular}{llccccl}
\hline
\noalign{\smallskip}  
\textbf{Parameter}         & \textbf{Visit 2}         & \textbf{Visit 3}                 & \textbf{Unit} \\                           
\noalign{\smallskip}
\hline
\hline
\noalign{\smallskip}
$\dot{M}_{\mathrm{H^{0}}}$                &     2.5$\stackrel{+1.1}{_{-0.8}}\times$10$^{8}$         & 2.5$\stackrel{+0.8}{_{-0.6}}\times$10$^{8}$  & g\,s$^{-1}$    \\ 
$\Gamma_{\mathrm{ion}}$                   & 2.2$\stackrel{+0.9}{_{-0.8}}\times$10$^{-5}$         &  2.4$\stackrel{+1.0}{_{-1.6}}\times$10$^{-5}$   & s$^{-1}$    \\ 
$v^{\mathrm{p}}_{\mathrm{wind}}$      &  50$\stackrel{+5}{_{-5}}$              & 60$\stackrel{+6}{_{-6}}$     & km\,s$^{-1}$\\
$V^{\mathrm{st}}_{\mathrm{bulk-wind}}$     & 85$\stackrel{+6}{_{-12}}$   &    85$\stackrel{+6}{_{-16}}$         &    km\,s$^{-1}$\\
$T^{\mathrm{st}}_{\mathrm{wind}}$         &  1.2$\pm$1.2$\times$10$^{4}$          &   1.2$\pm$1.2$\times$10$^{4}$     &  K \\
$v^{\mathrm{st}\dagger}_{\mathrm{therm-wind}}$         &   10$\pm$10          &   10$\pm$10     &  km\,s$^{-1}$\\
$n^{\mathrm{st}}_{\mathrm{wind}}$         &  1.3$\stackrel{+0.5}{_{-0.4}}\times$10$^{3}$       & 3.3$\stackrel{+1.5}{_{-1.0}}\times$10$^{3}$       & cm$^{-3}$   \\
\hline
$\chi^{2}$         &   203       &   182     &  \\
$dof$         &   200    &   198     &  \\
\hline                                                                  
\hline
\multicolumn{4}{l}{$\dagger$: The thermal velocity is calculated from the values obtained for the stellar wind kinetic temperature.} \\
\multicolumn{4}{l}{Note: $n^{\mathrm{st}}_{\mathrm{wind}}$, $\Gamma_{\mathrm{ion}}$, and $\dot{M}_{\mathrm{H^{0}}}$ are given at the distance of the semi-major axis ($a_{\mathrm{p}}$=0.0287\,au)} \\
\end{tabular}
\label{results_fits}
\end{table*}

\subsubsection{High-velocity stellar wind}
\label{high_wind}

Because the Lyman-$\alpha$ line extends between $\sim$~$\pm$200\,km\,s$^{-1}$, interactions between a stellar wind moving faster than this limit and the planetary exosphere will produce a different observational signature than in the first scenario (Sect.~\ref{sect:lowvel_results}). In that case, neutralized protons created by charge exchange have radial velocities that are too high to be visible in transmission in the Lyman-$\alpha$ line, and only their abrasion of planetary neutrals affects the observed signature (Sects.~\ref{sect:sec_pop} and \ref{sect:separation}). We investigated whether the observations of GJ\,436 b might be explained by interactions between the exosphere and a fast stellar wind.\\
To explore this scenario, $V^{\mathrm{st}}_{\mathrm{bulk-wind}}$ was arbitrarily fixed to a value higher than 200\,km\,s$^{-1}$. With the planetary neutral velocities below $\sim$120\,km\,s$^{-1}$ (\citealt{Bourrier2015_GJ436}), any value of the stellar wind bulk velocity beyond 200\,km\,s$^{-1}$ will produce the same results (because it will only influence the probability for a planetary neutral to be abraded; see Eq.~\ref{eq:ENA}), and it was fixed to 350\,km\,s$^{-1}$. The kinetic dispersion of the proton distribution has little influence on the charge-exchange probability as well and was fixed to $v^{\mathrm{st}}_{\mathrm{therm-wind}}$=20\,km\,s$^{-1}$. We then searched for the best fit by varying other free parameters. We were unable to find a good fit for Visit 3, as expected from the similar absorption signatures observed during both the transit and post-transit phases, which can only be produced by low-velocity neutralized protons in the exospheric tail (thus visible in transmission in the stellar line; Sect.~\ref{sect:sec_pop}). The best fit for Visit 2 was obtained for the following values: $\dot{M}_{\mathrm{H^{0}}}$=1.6$\times$10$^{9}$\,g\,s$^{-1}$, $\Gamma_{\mathrm{ion}}$=1.2$\times$10$^{-5}$\,s$^{-1}$, $v^{\mathrm{p}}_{\mathrm{wind}}$=45\,km\,s$^{-1}$, and $n^{\mathrm{st}}_{\mathrm{wind}}$=3.6$\times$10$^{3}$\,cm$^{-3}$, yielding a $\chi^{2}$ of 209.  \\
These values are similar to those obtained in the low wind-speed scenario (Sect.~\ref{sect:lowvel_results}), but the quality of the fit is lower for Visit 2 data, and a high-velocity wind scenario does not match Visit 3 data. This is in stark contrast with the low-velocity scenario that could explain both Visits 2 and 3 with very similar stellar wind and mass-loss properties, consistent with the stability of GJ\,436 in the Lyman-$\alpha$ line and in the X-rays (\citealt{Ehrenreich2015}; \citealt{Bourrier2015_GJ436}). Hereafter, we therefore assume that the most likely scenario for the observations of GJ\,436 b is a low-velocity stellar wind allowing for both abrasion and proton neutralization in the planetary exosphere. \\


\section{Interpretation of the results}
\label{interpret}

\subsection{Stellar winds and planetary exospheres}

In the models matching the observations, GJ\,436b is not in saturation regimes where either all escaping planetary neutrals interact with the stellar wind, or all stellar wind protons crossing the exosphere are neutralized by charge exchange. The observations of the warm Neptune constrain the exosphere to be composed of both planetary neutrals and neutralized protons. Planetary neutrals are mainly present in the coma close to the planet, while neutralized protons dominate the cometary tail (see Fig.~\ref{velfield_rad_stwind} and Sect.~\ref{struct}). This situation is different for the hot Jupiter HD\,189733 b, where the isolated absorption signature observed at very high velocity in the blue wing of the Lyman-$\alpha$ line was representative of a neutralized proton population alone (\citealt{Lecav2012}). Interactions of the K-type star HD189733 with its hot-Jupiter companion ($a$ = 0.031\,ua or 8.8\,R$_{*}$) are stronger than for the GJ436 system, with stellar wind velocities in excess of $\sim$200\,km\,s$^{-1}$ and a proton density in the range 4$\times$10$^{3}$ - 5$\times$10$^{7}$cm$^{-3}$ (\citealt{Bourrier_lecav2013}; \citealt{BJ_ballester2013}). By comparison, the warm Neptune GJ\,436 b ($a$ = 0.0287\,au or 14.1\,R$_{*}$) is subject  to wind velocities in the order of 70 - 90\,km\,s$^{-1}$ and proton densities in the range 1$\times$10$^{3}$ - 5$\times$10$^{3}$cm$^{-3}$ from its M dwarf host star. The properties of HD\,189733 stellar wind at the orbit the planet are most likely variable over time (see \citealt{Bourrier_lecav2013}; \citealt{Llama2013}; \citealt{Cauley_2015}, Fares et al. 2016 in prep.), and this also seems to be the case for GJ\,436 with a proton density in Visit 2 lower than in Visit 3. This may be linked to temporal variability or non-homogeneities in the stellar magnetic field at the orbital distance of the planet, which could be further investigated using spectropolarimetric observations of the M dwarf.\\

\subsection{Planetary mass loss}
\label{massloss}

\subsubsection{Energy-limited escape rate and ionization fraction}
\label{nrj_diag}

In the energy-limited regime, a fraction of the stellar X/EUV energy received by the upper planetary atmosphere is converted into mass loss and compensates for the gravitational potential energy required by the gas to escape. Based on results of \citet{Owen2016} (see their Figures 1 and 4), we infer that GJ\,436 b is located in this regime, since the warm Neptune has a mass of 1.4$\times$10$^{29}$\,g (\citealt{Butler2004}) and radius of 2.5$\times$10$^{9}$\,cm (\citealt{Knutson2011}), and orbits a star with EUV emission in the order of 10$^{2}$\,erg\,s$^{-1}$\,cm$^{-2}$ at the semi-major axis\footnote{For the purpose of comparison with \citet{Owen2016}, we calculated this value accounting from photons above 40\,nm, as given in Table~\ref{XEUV_flux}.}. The escape rate of neutral hydrogen can therefore be expressed as (e.g., \citealt{Bourrier2015})
\begin{equation}
\label{eq:H_esc_rate}
\dot{M}^{tot}= \eta \, \frac{3 \, F_\mathrm{X/EUV}(\mathrm{1 au})}{4 \, G \, a_\mathrm{p}^{2} \, \overline{\rho} \, K_{tide}},\end{equation}
with $\overline{\rho}$ the mean density of the planet, $K_\mathrm{tide}$ a correction factor accounting for the contribution of tidal forces to the potential energy (\citealt{Erkaev2007}), and $\eta$ the heating efficiency, which the most recent theoretical estimations estimate at between 10 and 20\% (e.g., \citealt{Lammer2013}; \citealt{Shematovich2014}; \citealt{Owen2016}). The total X/EUV flux per unit area at 1\,au from GJ\,436, measured from 0.5\,nm to the Lyman-$\alpha$ line (included), is $F_\mathrm{X/EUV}(\mathrm{1 au}) = $2.3$\pm$0.5\,erg\,s$^{-1}$\,cm$^{-2}$ (Table~\ref{XEUV_flux}). The corresponding total mass-loss rate from GJ\,436b atmosphere is $\dot{M}^{tot} = \eta\,\dot{M}^{100\%}$, with $\dot{M}^{100\%} = $2.2$\pm$0.6$\times$10$^{10}$\,g/s. Defining $\dot{M}_\mathrm{H^{0}}=f_\mathrm{H^{0}}\,\dot{M}^{tot}$, where $f_\mathrm{H^{0}}$ is the neutral fraction of hydrogen in the upper thermosphere of the planet and $\dot{M}_\mathrm{H^{0}}$ the escape rate of neutral hydrogen obtained with EVE, we found that $\eta\,f_\mathrm{H^{0}}$ should be 1.2$\pm$0.5$\times$10$^{-2}$. \\

\subsubsection{Magnetically driven outflow}
\label{mag_outflow}

\begin{figure}  
\includegraphics[trim=0cm 0cm 0cm 2.5cm,clip=true,width=\columnwidth]{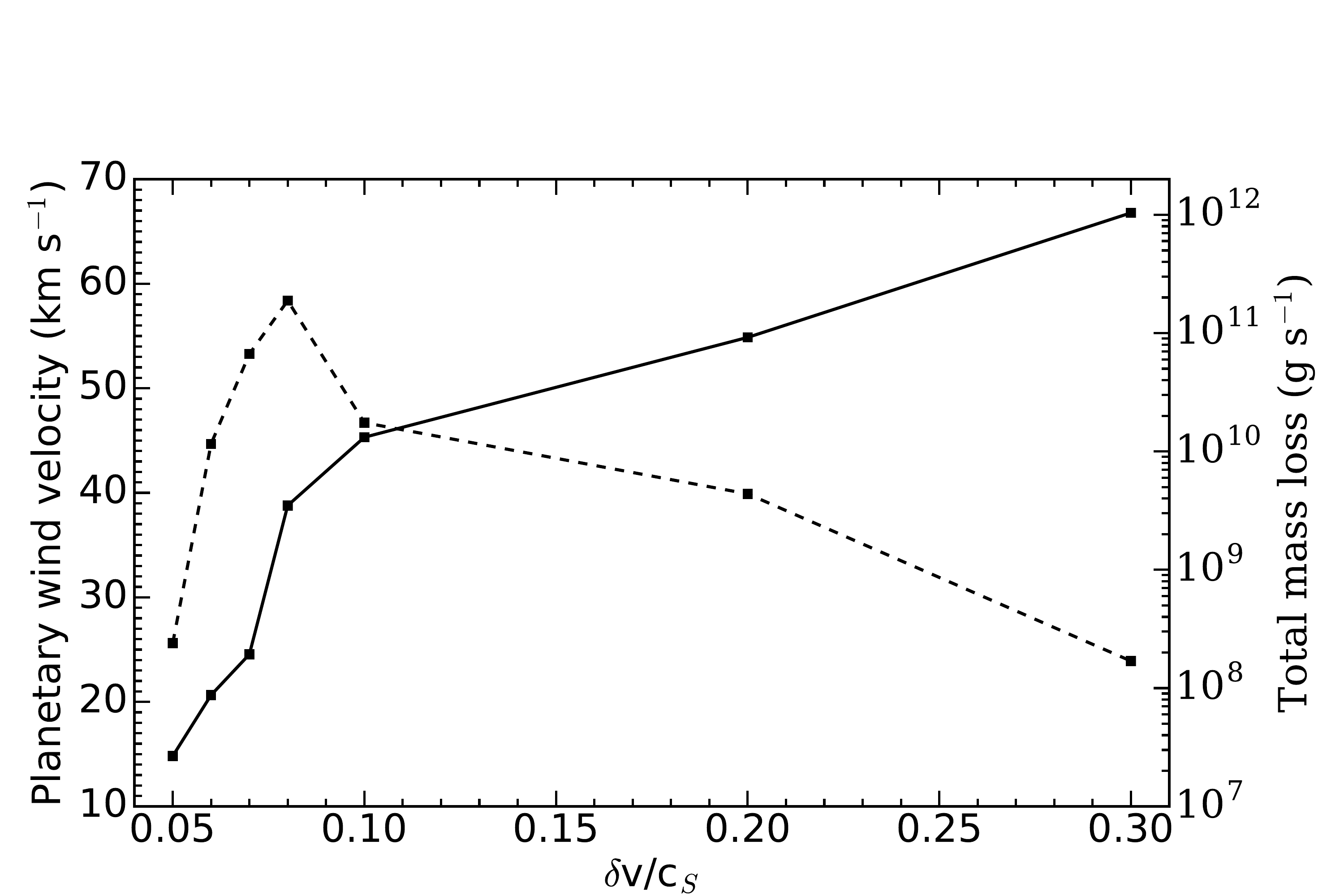}
\caption[]{Dependence of the maximum wind velocity (dashed line) and total planetary mass-loss rate (solid line) at the Roche radius as a function of the velocity dispersion at the planet surface (note that the two vertical axes do not correspond). Results come from 1D simulations of atmospheric outflows induced by MHD waves. Ranges of $v^{\mathrm{p}}_{\mathrm{wind}}$ values derived from the observations are highlighted as blue (Visit 2) and red (Visit 3) shaded areas, allowing an estimation of the required velocity dispersions and corresponding total mass loss.}
\label{mag_driven_flow}
\end{figure}

Our simulations show that the atmospheric escape velocity must be in the range 40 - 70\,km\,s$^{-1}$ to explain the size of the exosphere (Table~\ref{results_fits}). This outflow velocity is faster than predicted for hot Jupiters ($\sim$1-10\,km\,s$^{-1}$; see, e.g., \citealt{MurrayClay2009}; \citealt{Koskinen2013a}). To solve this discrepancy, we investigated the possibility that the outflow might arise from MHD waves in the upper atmosphere of GJ\,436b. In the presence of a magnetic field, the turbulence of the gas can excite MHD waves that dissipate in the upper atmosphere and drive the outflow (\citealt{Suzuki_Inutsuka_2005,Suzuki_Inutsuka_2006}). We used the model described in \citet{Tanaka2014} to calculate the 1D atmospheric structure of GJ\,436 b for different values of the velocity dispersion at the surface. The velocity that we observed at the Roche lobe can be obtained with dispersions of about [6 - 17]\% of the sound speed (2.4\,km\,s$^{-1}$ at the planet surface, assuming a temperature of 800\,K). Under these conditions, the surface velocity dispersion is high enough to drive a fast planetary wind, but low enough that the increasing density of the outflow does not hinder its acceleration (Fig.~\ref{mag_driven_flow}). Because of the nonlinear dissipation of the MHD wave energy in the upper atmosphere, the structure of the outflow is time variable (\citealt{Tanaka2015}), and the values measured here for $v^{\mathrm{p}}_{\mathrm{wind}}$ are reached when the outflow is close to its maximum speeds. Theoretical studies have shown that Alfvenic waves can be severely damped at low altitudes in the atmosphere, confining the heating and enhanced mass loss in specific regions such as the magnetic poles (\citealt{Trammell2011}; \citealt{Tanaka2015}; \citealt{Khodachenko2015}). For GJ\,436b the large measured transit depth requires a giant coma that surrounds the planet on all sides, which in turn requires gas to escape from equatorial as well as polar regions. We conclude that the planetary outflow is unlikely to be confined to a specific region of the upper atmosphere but can be magnetically driven by MHD waves that dissipate in the upper atmosphere, which would
explain the high velocity observed at the base of GJ\,436 b exosphere. A self-consistent 3D model including both a MHD-described thermosphere and particle-described exosphere is nonetheless required to fully explore this scenario.\\

Moreover, when the time-variable outflow is close to its maximum speed, simulations of the magnetically driven outflow yield a total atmospheric mass-loss rate $\dot{M}^{mag}$ in the range [8$\times$10$^{7}$ , 5$\times$10$^{10}$]\,g/s, which agrees remarkably well with the observations. The comparison between this calculated total escape rate $\dot{M}^{mag}$, the total escape rate in the energy limited regime $\dot{M}^{tot}$, and the observed escape rate of neutral hydrogen $\dot{M}_\mathrm{H^{0}}$ allows us to further constrain the heating efficiency $\eta$ and neutral fraction of hydrogen $f_\mathrm{H^{0}}$. Using the relations $\dot{M}^{mag} = \dot{M}^{tot} = \eta\,\dot{M}^{100\%} = \dot{M}_\mathrm{H^{0}} / f_\mathrm{H^{0}}$, the estimated $\dot{M}^{100\%}$ (Sect.~\ref{nrj_diag}) and the measured $\mathrm{H^{0}}$ (Table~\ref{results_fits}), we find lower limits of $\sim$0.5\% for both $\eta$ and $f_\mathrm{H^{0}}$. This is consistent with the value of about 1\% for $\eta$ estimated by \citet{Ehrenreich2015}.\\

In conclusion, a magnetically driven outflow provides a consistent scenario for the observed escaping planetary wind, in which the measured values for the wind velocity and the mass-loss rate (and the related heating efficiency and neutral fraction) are well explained.

\subsection{Planetary magnetic moment}

Magnetic interactions between close-in planets and their host star may play a part in shaping planetary exospheres (e.g., \citealt{Vidotto2011a}; \citealt{Matsakos2015}). Analysis of radio emission from an exoplanet would constrain the planetary magnetic field strength, but there has been no confirmed detection to date (e.g., \citealt{Bastian2000}; \citealt{Lazio2004}; \citealt{Smith2009}; \citealt{Hallinan2013}, but see also \citealt{Lecav2013}, \citealt{Sirothia2014}). Because the charged stellar wind protons can interact with the exosphere of an exoplanet and be deflected by its magnetosphere, observations in the Lyman-$\alpha$ line can potentially be used to estimate the planetary magnetic field in addition to the stellar wind properties. In their study of the hot Jupiter HD209458b, this approach was chosen by \citet{Holmstrom2008} and \citet{Ekenback2010}, who prescribed a magnetic obstacle surrounding the planet that prevented penetration by protons, and by \citet{Kislyakova2014}, who estimated the magnetic moment of the planet. In that case, however, the uncertainties and phase coverage of the Lyman-$\alpha$ observations lead to strong degeneracies between the relative contribution of the planetary escape, stellar wind, and magnetic obstacle properties, and a scenario with radiation pressure alone was even found to explain the observations well (\citealt{VM2003}; \citealt{Lecav2008}; \citealt{Bourrier_lecav2013}). Similar conclusions were reached by \citet{BJ_ballester2013} when adjusting the observations of ionized carbon around HD\,189733b, with many possible solutions for the stellar wind and planetary magnetosphere properties. \\
We showed for GJ\,436b that the Lyman-$\alpha$ transit observations are well explained by the combination of radiative braking and stellar wind interactions. Therefore we did not include a magnetic obstacle in the EVE simulations because it would bring unnecessary additional free parameters to the interpretation of the data. It is nonetheless possible to use our simulation results to set an upper limit on the magnetic moment of GJ\,436b. Consider that the planet is surrounded by a magnetosphere, which would prevent stellar wind protons from interacting with the planetary exosphere. Since our best-fit simulations show significant stellar wind interactions down to about 20\,$R_{p}$ from the planet at the time of the transit (see Figs.~\ref{velfield_rad_stwind} and~\ref{dyn_time}), this corresponds to the largest possible stand-off distance $R_s$ of the magnetosphere. At this altitude, a temperature of several 10$^{6}$\,K would be needed for the thermal pressure from the planetary wind to overcome the stellar wind forcing. Therefore we consider that the contribution from thermal pressure is negligible and that the magnetospheric size is set by the balance between the planetary magnetic pressure and the stellar wind ram pressure, yielding a planetary magnetic moment (e.g., \citealt{griessmeir2004}; \citealt{See2014})
\begin{equation}\label{eq.magmoment}
\mathcal{M}=\left(\frac{8 \pi^2 R_{s}^{6} \rho_{\mathrm{wind}} v_{\mathrm{rel}}^{2} }{\mu_0 f_0^2}\right)^{1/2}, 
\end{equation} 
where we assumed the planet to have a dipolar field, $v_{\mathrm{rel}}$ is the velocity of the stellar wind relative to the planet, $\mu_{0}$ the permeability of vacuum, and $f_{0}\approx$1.22 is a form factor of the magnetosphere. We set the mass density of the stellar wind at the planetary orbit, $\rho_{\mathrm{wind}}\simeq (2 - 5)\times10^{-18}$\,kg\,m$^{-3}$, to its average value at the center of the transit for Visits 2 and 3 (Table~\ref{results_fits}). At this time, the planet is close to its semi-minor axis, and its orbital velocity is nearly perpendicular to the radial bulk velocity of the stellar wind (\citealt{Vidotto2011b}), with $v_{\mathrm{rel}}$ = $\sqrt{v_{\mathrm{p}}^2 + v_{\mathrm{bulk}}^2} \sim$ 140\,km\,s$^{-1}$. From Equation (\ref{eq.magmoment}), we find
\begin{equation}
\mathcal{M} \lesssim 2.5 \times 10^{26} \mathrm{\,A\,m}^{2} \sim 0.16\,\mathcal{M}_\mathrm{Jup},
\end{equation} 
where $\mathcal{M}_\mathrm{Jup}=1.56\times10^{27}${\,A\,m}$^{2}$ is the magnetic moment of Jupiter (\citealt{griessmeir2004}). For a planetary radius $R_p = 0.64 R_\mathrm{Jup}$, this implies an upper limit for the equatorial magnetic field strength of GJ\,436b of
\begin{equation}
\frac{B_p}{B_{\rm Jup}} =  \frac{\mathcal{M}}{\mathcal{M}_\mathrm{Jup}} \left(\frac{R_p}{R_\mathrm{Jup}}\right)^{-3} \lesssim \, 0.6,
\end{equation} 
or $B_p \lesssim 2.7$G, assuming an equatorial magnetic field strength of  ${B_{\rm Jup}} \sim 4.3$G for Jupiter (\citealt{Bagenal2013}). This is consistent with the magnetic field strength of $\sim$1\,G required to accelerate a magnetically driven outflow to the observed velocities (Sect.~\ref{mag_outflow}). We caution that the magnetic field strength we derive here is a conservative upper limit, and, at the strength of 1\,G, a self-consistent 3D model would be required to explore the possible escape anisotropies. The magnetosphere of the planet is likely to lie well inside the 20\,$R_{p}$ estimate for the stand-off distance, as our analysis shows that stellar wind particles interact with the (unprotected) planetary exosphere at such distances.\\


\section{Discussion: separating radiation pressure and stellar wind contributions}
\label{sect:separation}

We discuss in this section the general possibility of distinguishing the contributions of radiation pressure and stellar wind interactions when analyzing Lyman-$\alpha$ line observations of an evaporating exoplanet exosphere. \\
\begin{itemize}
\item The most direct case to identify the contribution of the stellar wind is when its projected velocity is lower than the Doppler width of the Lyman-$\alpha$ line, but beyond the highest velocity that can be reached under radiation pressure acceleration. An absorption signature beyond this limit can then be attributed to a population of neutralized protons independent of the planetary neutrals, as was the case for the hot Jupiter HD\,189733b (\citealt{Lecav2012}). Interestingly, no other Lyman-$\alpha$ absorption signature was detected for this planet at radiation-pressure-induced velocities, presumably because of high photo-ionization rates or massive stellar wind abrasion of the escaping planetary neutrals (\citealt{Bourrier_lecav2013}).\\
\item If interacting stellar wind protons have projected velocities higher than the Doppler width of the Lyman-$\alpha$ line, they can still be detected through their abrading effect. However, the phase coverage of the transit must then allow very different regions of the exosphere to be probed. For the hot Jupiter HD\,209458b, observations after the optical transit have been obtained only once at low signal-to-noise ratio (\citealt{Ehrenreich2008}), and therefore available data can be interpreted either by radiative blow-out alone (\citealt{Bourrier_lecav2013}) or by the addition of a ~$\sim$400\,km\,s$^{-1}$ stellar wind with higher escape rates (\citealt{Holmstrom2008}, \citealt{Ekenback2010}, \citealt{Kislyakova2014}). In that case, the decrease in absorption depth caused by the fast wind abrasion can indeed be compensated for by variations in other parameters such as the planetary escape rate or stellar photoionization rate, leading to degeneracies in the stellar wind and planetary escape properties. \\
\item When stellar wind and radiatively induced projected velocities overlap, the observed absorption results from a balance between planetary neutral hydrogen atoms and neutralized stellar wind protons. As in the previous case, different combinations of parameters such as proton density and planetary escape rate can yield similar absorption profiles at a given time, but \citet{Bourrier_lecav2013} proposed that the two mechanisms may be distinguished by analyzing the spectro-temporal variations of the absorption profile. For radiation-pressure-driven mechanisms, a radiative blow-out creates a narrow cometary tail with a strong velocity gradient (e.g., for HD\,209458b; \citealt{Bourrier_lecav2013}), whereas radiative braking leads to a massive expansion of the exosphere and its dilution within a broad cometary tail (\citealt{Bourrier2015_GJ436}; see also Fig.~\ref{spectra}). In both cases, the depth of the absorption at a given wavelength varies strongly over time during the transit. In contrast, for the stellar-wind-driven mechanism, we showed in Sect.~\ref{sect:sec_pop} that a population of neutralized stellar wind protons is characterized by a time-stable repartition of the absorption depth with wavelength. These spectro-temporal variations of the absorption profile could not be analyzed for HD\,209458b because of the reduced phase coverage and were only tentatively studied for HD\,189733 b (\citealt{Bourrier2013}). \\
With a good phase coverage of the exospheric transit and very large absorption depths that magnify the spectro-temporal variations of the absorption, the observations of GJ\,436 b allow us, for the first time, to clearly separate the contributions of radiation pressure and stellar wind and to probe the regions of the exosphere that are shaped by each mechanism. 
\end{itemize}


\section{Conclusion}
\label{sect:conclu}

We investigated the physical conditions in the exosphere of the warm Neptune GJ\,436b observed at three different epochs in the Lyman-$\alpha$ line. We independently interpreted the spectra at each epoch through direct comparison with 3D numerical simulations performed with the code EVE. While radiative braking was shown to play a major role in shaping the exosphere of GJ\,436b, specific features remained to be explained in the observations, such as sharp egresses in Visits 2 and 3. \\
To this aim, we studied the additional effect of the stellar wind and found that significant interactions with the exosphere are required to explain the observations. With low velocities in the order of $\sim$85\,km\,s$^{-1}$, the stellar wind both abrades the planetary exosphere and leads to the formation of a secondary population of neutralized protons that contribute to the observed absorption in the blue wing of the Lyman-$\alpha$ line. The combination of radiation pressure and stellar wind abrasion allows for the formation of a giant coma with a reduced front ahead of the planet that reproduces the observed early ingresses. Furthermore, the different dynamics of the neutralized proton and planetary neutral populations allow for the formation of a cometary tail that either produces a sharp (Visit 2) or a flat egress (Visit 3), depending on the relative balance between the two populations. This balance seems to be determined mainly by changes in the density of the stellar wind between the two epochs (from $\sim$10$^{3}$\,cm$^{-3}$ in Visit 2 to 3$\times$10$^{3}$\,cm$^{-3}$ in Visit 3) and a marginally lower planetary outflow velocity in the second epoch ($\sim$50\,km\,s$^{-1}$ against 60\,km\,s$^{-1}$), while the other physical conditions in the upper atmosphere of GJ\,436 b are otherwise very stable for the planetary mass loss ($\sim$2.5$\times$10$^{8}$\,g\,s$^{-1}$), hydrogen photoionization rate ($\sim$2$\times$10$^{-5}$\,s$^{-1}$), and stellar wind bulk ($\sim$85\,km\,s$^{-1}$) and thermal velocities ($\sim$10\,km\,s$^{-1}$). \\
Using EVE simulations, we detailed how these properties influence the structure of the exosphere and compared the best-fit results with independent theoretical estimations. Comparisons with energy-limited escape rates place constraints on the heating efficiency $\eta$ of the upper atmosphere and its neutral hydrogen content $f_\mathrm{H^{0}}$, with $\eta\,f_\mathrm{H^{0}}\sim$10$^{-2}$ and a lower limit on both $\eta$ and $f_\mathrm{H^{0}}$ of about 0.5\%. The properties of the stellar wind for the low-mass star GJ\,436 are not unexpected, with lower density and bulk velocity than for earlier host stars, but it is the first time that these properties are directly measured from observations for a M dwarf. Future observations of extended atmospheres in similar systems therefore have a high potential for stellar wind characterization. On the other hand, the velocity of the escaping gas is faster than the outflows of evaporating hot Jupiters. The shallower gravity well of the Neptune-mass GJ\,436 b may play some part, although its thermosphere is less irradiated. We showed that a possible mechanism for the fast ouflows might be turbulence-driven MHD waves at the surface of the planet. While the planetary magnetic field cannot be strongly constrained by Lyman-$\alpha$ observations, we used a geometric argument to set an upper limit on GJ\,436b of about a tenth of Jupiter magnetic moment. \\
We note that Visit 2 observations might also be explained through pure abrasion of the exosphere from a high-velocity stellar wind. However, this scenario involves escape rates expected from more strongly irradiated hot Jupiters and is not consistent with Visit 1 or Visit 3 observations. Given the stability of GJ\,436 stellar Lyman-$\alpha$ line in four epochs covering four years of observations (see \citealt{Ehrenreich2015}) and the fact that Visits 2 and 3 can be explained with very similar conditions for the stellar irradiation, the stellar wind, and planetary mass-loss properties, this second scenario seems very unlikely. We also note that we were unable to adjust Visit 1 spectra with the model used in this paper, most probably because of the lack of reference at this epoch for the intrinsic stellar Lyman-$\alpha$ line, but also because the observations hint at more significant variations in the physical conditions of the exosphere at this epoch. \\
We emphasize that future models of the GJ\,436b exosphere should make use of the entire spectral content of the Lyman-$\alpha$ observations, distributed over 24 exposures and more than 600 data points. The exploration of the six-parameter space for the three epochs of observations and the different scenarii investigated required about 18000 simulations, running full-time for nearly a year on 15 dual-processors compute nodes totalling 276 cores. EVE simulations show that the neutralized protons populating the exosphere would be photoionized much farther from GJ\,436b than planetary neutrals, making the transit of the cometary tail visible at Lyman-$\alpha$ for more than half the revolution period of the planet. New observations covering later phases than previously observed will allow for a full caracterization of the shape and properties of the exosphere, refining the measurements of the stellar wind properties.\\


\begin{acknowledgements}
This work is based on observations made with the NASA/ESA Hubble Space Telescope, obtained at the Space Telescope Science Institute, which is operated by the Association of Universities for Research in Astronomy, Inc., under NASA contract NAS 5-26555. This work has been carried out in the frame of the National Centre for Competence in Research ``PlanetS'' supported by the Swiss National Science Foundation (SNSF). The authors wish to thank the referee for their useful report. V.B., D.E., and A.A.V acknowledge the financial support of the SNSF. A.L.E acknowledges financial support from the Centre National d\textsc{\char13}Etudes Spatiales (CNES). The authors acknowledge the support of the French Agence Nationale de la Recherche (ANR), under program ANR-12-BS05-0012 ``Exo-Atmos''. Y.A.T thanks T. K. Suzuki for developing a simulation code for magnetically driven outflow and fruitful discussions.  
\end{acknowledgements}

\bibliographystyle{aa} 
\bibliography{biblio} 

\end{document}